\documentclass[a4paper,11pt,nohyper]{JHEP3}
\def\a{\alpha}
\def\b{\beta}

\def\d{\delta}

\def\l{\lambda}

\def\t{\tau}

\def\S{\Sigma}

\def\del{\partial}              
\let\a=\alpha \let\b=\beta  \let\d=\delta 
    
\let\l=\lambda    
 \let\t=\tau

\def\nn{\nonumber} \def\bd{\begin{document}} \def\ed{\end{document}}
\def\ds{\documentstyle} \let\fr=\frac \let\bl=\bigl \let\br=\bigr
\let\Br=\Bigr \let\Bl=\Bigl
\let\bm=\bibitem
\let\na=\nabla
\let\pa=\partial \let\ov=\overline
\newcommand{\be}{\begin{equation}}
\newcommand{\ee}{\end{equation}}
\def\ba{\begin{array}}
\def\ea{\end{array}}
\def\ft#1#2{{\textstyle{{\scriptstyle #1}\over {\scriptstyle #2}}}}
\def\fft#1#2{{#1 \over #2}}
\def\del{\partial}
\def\sst#1{{\scriptscriptstyle #1}}
 \def\oneone{\rlap 1\mkern4mu{\rm l}}
\def\ie{{\it i.e.\ }}
\def\via{{\it via}}
\def\semi{{\ltimes}}
\def\str{{\rm str}}
\def\Dm{{{D_{\sst{max}}}}}
\def\vac{ \left | 0 \right \rangle }
\def\kvac{ \left | k \right \rangle }

\def\sp{\; \; \;}

\def\bol{ \left | B (p^+) \right \rangle}
\def\bo1{ \left | B^0 (p^+) \right \rangle}

\def\bolt{ \left | B (p^+) \right \rangle_{\t}}

\def\boxl{ \left | B (x^-) \right \rangle}
\newcommand{\bea}{\begin{eqnarray}}
\newcommand{\eea}{\end{eqnarray}}

\def\<{ \langle }
\def\>{ \rangle }

\def\S{\Sigma}

\usepackage{amsmath,latexsym,amssymb,slashed}
\usepackage{graphicx}
\usepackage[latin1]{inputenc}
\usepackage{subfigure}

\renewcommand{\floatpagefraction}{0.6}
\renewcommand{\textfraction}{0.2}

\newcommand\ca{\mathcal{A}}
\newcommand\vp{\varphi}
\newcommand\beal{\begin{align}}
\newcommand\bbone{\ensuremath{\mathbbm{1}}}

\newcommand{\eq}[1]{\begin{equation}#1\end{equation}}
\newcommand{\spl}[1]{\begin{split}#1\end{split}}
\newcommand{\al}[1]{\begin{align}#1\end{align}}
\newcommand{\subeq}[1]{\begin{subequations}#1\end{subequations}}

\newcommand{\arXividhepth}[1]{\href{http://arxiv.org/abs/#1}arXiv:{\tt #1} [hep-th]}
\newcommand{\arXividother}[2]{\href{http://arxiv.org/abs/#1}arXiv:{\tt #1} [#2]}

\newcommand{\bg}[1]{\hat{#1}}
\newcommand{\wj}{\widetilde{J}}
\newcommand{\reo}{\mathrm{Re}~\!\omega}
\newcommand{\imo}{\mathrm{Im}~\!\omega}
\newcommand{\ads}{AdS_4}
\newcommand{\mcal}{\mathcal{M}}
\newcommand{\ccal}{\mathcal{C}}
\newcommand{\ncal}{\mathcal{N}}
\newcommand{\boxedeq}[1]{
\begin{equation}
\fbox{
\rule[0.7cm]{0pt}{0pt}
$#1$
\rule[-0.45cm]{0pt}{0pt}
}
\end{equation}
}

\def\d{\text{d}}

\def\slashchar#1{\setbox0=\hbox{$#1$}           
\dimen0=\wd0                                 
\setbox1=\hbox{/} \dimen1=\wd1               
\ifdim\dimen0>\dimen1                        
\rlap{\hbox to \dimen0{\hfil/\hfil}}      
#1                                        
\else                                        
\rlap{\hbox to \dimen1{\hfil$#1$\hfil}}   
/                                         
\fi}

\def\Re           {{\rm Re\hskip0.1em}}
\def\Im           {{\rm Im\hskip0.1em}}

\newcommand{\E}{\text{\tiny E}}

\newcommand{\tV}{{\widetilde{V}}}
\newcommand{\tH}{{\tilde{h}}}
\newcommand{\tm}{{{m}}}
\newcommand{\tmu}{{\tilde{\mu}}}
\newcommand{\trho}{{\tilde{\rho}}}
\newcommand{\tv}{{\tilde{v}}}
\newcommand{\calo}{\mbox{${\cal O}$}}
\newcommand{\cala}{\mbox{${\cal A}$}}
\newcommand{\dd}{\mathrm{d}}
\newcommand{\ra}{\rightarrow}
\newcommand{\calv}{\mbox{${\cal V}$}}
\newcommand{\calh}{\mbox{${\cal H}$}}
\newcommand{\calm}{\mbox{${\cal M}$}}
\newcommand{\abs}[1]{\left| #1 \right|}
\newcommand{\zetaa}{{\psi}}
\newcommand{\tr}{{\rm tr}\,}

\newcommand{\ky}[1]{{\color{blue}{#1}}}

\title{Generalized entanglement entropy}

\author{Marika Taylor${}^{\diamondsuit}$  \\

\begin{itemize}
 \renewcommand{\labelitemi}{${}^\diamondsuit$}
\item School of Mathematical Sciences, University of Southampton, \\
Highfield, Southampton, SO17 1BJ, UK.

  \end{itemize}

\bigskip
 E-mail:
 \email{m.m.taylor@soton.ac.uk}}

\abstract{We discuss two measures of entanglement in quantum field theory and their holographic realizations. For field theories admitting a global symmetry, we introduce a global symmetry entanglement entropy, associated with the partitioning of the symmetry group. This quantity is proposed to be related to the generalized holographic entanglement entropy defined via the partitioning of the internal space of the bulk geometry. The second measure of quantum field theory entanglement is the field space entanglement entropy, obtained by integrating out a subset of the quantum fields. We argue that field space entanglement entropy cannot be precisely realised geometrically in a holographic dual. However, for holographic geometries with interior decoupling regions, the differential entropy provides a close analogue to the field space entanglement entropy. We derive generic descriptions of such inner throat regions 
in terms of gravity coupled to massive scalars and show how the differential entropy in the throat captures features of the field space entanglement entropy.}

\begin{document}

\newcommand{\td}{\tilde}
 \newcommand{\bc}{\begin{center}}
 \newcommand{\ec}{\end{center}}
 \newcommand{\bfr}{\begin{flushright}}
 \newcommand{\efr}{\end{flushright}}
 \newcommand{\bfl}{\begin{flushleft}}
 \newcommand{\efl}{\end{flushleft}}
 \newcommand{\bt}{\begin{tabular}}
 \newcommand{\et}{\end{tabular}}

\section{Introduction} \label{one}

The Ryu-Takayanagi formula \cite{Ryu:2006bv,Ryu:2006ef} for computing the entanglement entropy holographically has stimulated a huge amount of interest in studying quantum entanglement and its relation to gravity, see the review \cite{Takayanagi:2012kg}. One of the main goals is to understand how quantum entanglement captures global structure in the holographically dual spacetime and whether the latter can be reconstructed from entanglement. Ideas about spacetime reconstruction using entanglement can be found in \cite{Swingle:2009bg,VanRaamsdonk:2009ar,VanRaamsdonk:2010pw,Czech:2012bh,Bianchi:2012ev}.

In particular, it was proposed in \cite{Bianchi:2012ev} that the area of a non-minimal closed surface in a holographic geometry should be related to the entanglement between the degrees of freedom contained within this region and those of its complement. Recently a sharp relation between the area of such a hole and the differential entropy was shown, see \cite{Balasubramanian:2013rqa,Balasubramanian:2013lsa,Myers:2014jia,Czech:2014wka,Balasubramanian:2014sra,Headrick:2014eia} and section \ref{six-de}. The interpretation of the differential entropy in quantum information theory was further discussed in \cite{Czech:2014tva,Czech:2015qta} and limitations on spacetime reconstruction (``shadows") were discussed in \cite{Freivogel:2014lja,Engelhardt:2015dta}, see also \cite{Engelhardt:2013tra}. 

To develop our understanding of spacetime reconstruction, it would be desirable to extend the holographic realisation of the Ryu-Takayanagi formula to other measures of quantum entanglement. However,  many standard measures of quantum entanglement unfortunately do not seem to admit a simple holographic description. For example, the logarithmic negativity measures the distillable entanglement contained in a quantum state. This quantity has been explored in a number of recent papers including \cite{Rangamani:2014ywa,Kulaxizi:2014nma,Perlmutter:2015vma} but there is as yet no proposal for the holographic computation of the negativity. The negativity is known to be related to the Renyi entropy at index one half 
which would seem to suggest analytic extension to a non-integral number of copies of the bulk geometry might be necessary to realise the negativity holographically. 

In this work we discuss two measures of entanglement in quantum field theory and their holographic realisation. The first measure of entanglement corresponds to integrating out a subset of fields in the quantum field theory; we denote this the field space entanglement entropy, as it is associated with a partitioning of the field space. This quantity has previously been discussed in \cite{Furukawa:2010nd,Xu:2011gn,Chen:2013kba,Lundgren:2013gba,Mollabashi:2014qfa}. The second measure of entanglement is applicable only to field theories with a global 
symmetry. It corresponds to integrating out part of the orbit of the global symmetry and we hence denote it as the global 
symmetry entanglement entropy. 

In sections \ref{two} and \ref{three} we discuss features of these entanglement entropies in simple field theory models. For both quantities the leading UV divergences scale with the spatial volume, as one would expect, since entanglement with the modes which have been integrated out occurs throughout the spatial region. Consider the symmetry preserving vacuum state in a conformal field theory with global symmetry, such as ${\cal N} = 4$ SYM. The global symmetry entanglement entropy is non-zero, and depends on how one partitions the global symmetry. To define the field space entanglement entropy one would have to integrate out some of the SYM fields; we can implement this perturbatively in quantum field theory but there seems to be no natural way to realise this situation holographically. 

Next consider a field theory which does not  have global symmetry. The global symmetry entanglement entropy can therefore clearly not be defined but let us suppose we can integrate out (massive) fields to define a field space entanglement entropy. In general the effective description after integrating out such modes may be expressed in terms of irrelevant operator deformations of a low energy action; this is the picture we should have in mind when trying to realise field space entanglement entropy holographically. 

\bigskip

Let us now turn to holographic analogues of these entropies. The global symmetry entanglement entropy is argued in section \ref{three} to correspond to the generalised holographic entanglement entropy introduced in \cite{Mollabashi:2014qfa} and discussed further in \cite{Karch:2014pma}. The latter is computed from the area of a codimension two minimal surface for which the boundary condition at conformal infinity is such that it fills the spatial background for the dual field theory and partitions the compact part of the geometry. That is, for any asymptotically $AdS \times S$ geometry the boundary condition for the minimal surface is a partitioning of the sphere. This partitioning is argued to be exactly the partitioning of the global symmetry orbit used in defining the global symmetry entanglement entropy. 

The field space entanglement entropy is more subtle as in general we would not expect that this quantity can be realised holographically. The field space entanglement entropy requires integrating out a subset of quantum fields but the holographic duals of the latter cannot in general be viewed as localised in the dual geometry. In the example given above, integrating out fields in the trivial vacuum of ${\cal N} = 4$ SYM, we do not expect a simple geometric realisation. 

The closest holographic analogues to the setup for field space entanglement entropy are situations in which the bulk geometry has interior throat regions, at which low energy degrees of freedom from the field theory are localised.  Whenever there is an inner throat region, there will be a dual description of this inner throat in terms of a quantum field theory with irrelevant deformations, which we argued above was the setup needed for field space entanglement entropy.

In sections \ref{d3} and \ref{sec:five} we consider Coulomb branch supergravity solutions for separated D3-brane, M2-brane, M5-brane and D1-D5 brane stacks. In such cases the geometries contain inner throat regions associated with each brane stack. The dual field theory description for each inner throat is a conformal field theory whose gauge group has a rank corresponding to the number of branes in the stack. As pointed out in \cite{Mollabashi:2014qfa} such geometries geometrically realise an analogue to the setup of field space entanglement entropy, as  within the inner throat regions we can view the degrees of freedom associated with the other brane stacks as having been integrated out. 

We use the methods of Kaluza-Klein holography \cite{Skenderis:2006uy,Skenderis:2006di,Skenderis:2007yb} in sections \ref{d3} and \ref{sec:five} to show that there is an effective low energy description of each inner throat region in terms of Einstein gravity coupled to massive scalar fields (dual to irrelevant operators in the conformal field theory associated with the throat). These massive scalar fields characterise geometrically the effect of integrating out the degrees of freedom associated with the other brane stacks. Note that this effective description is obtained by reducing over the sphere using \cite{Skenderis:2006uy,Skenderis:2006di}, and thus the holographic description has only one extra radial dimension relative to the field theory description. 

The definition of the field space entanglement entropy does not rely on the existence of any global symmetry. From the bulk perspective this implies that the compact part of the geometry should not be a prerequisite to describe the field space entanglement entropy; the only prerequisite should be an inner decoupling region. Our effective actions for the Coulomb branch geometries (after reducing over the sphere) indeed give exactly such descriptions of inner throat regions. In section \ref{sec:five} we consider other examples of  holographic geometries with interior decoupling regions: near extremal AdS Reissner-Nordstr\"{o}m black holes. We show that again the inner throats can be described by  gravity coupled to massive scalar fields. 

The effective holographic descriptions for interior throat regions are then used in section \ref{sec:six} to explore geometric measures of entanglement. We show that the area of a spatial hole in an inner throat is equivalent to the differential entropy, i.e. the analysis of \cite{Balasubramanian:2013rqa,Balasubramanian:2013lsa,Myers:2014jia,Czech:2014wka,Balasubramanian:2014sra,Headrick:2014eia} extends to the case in which one makes irrelevant deformations of the underlying conformal field theory. 

Both the spatial volume of the throat and the differential entropy capture features of the field space entanglement entropy. However, these geometric quantities are generically non-zero even for asymptotically AdS throats and are therefore not equivalent to the field space entanglement entropy. This was to be expected: by zooming into the inner throat region and imposing a cutoff there, we are effectively removing high energy modes from the low energy field theory dual to the throat itself. Thus the geometric measures of entanglement capture not just the entanglement with the degrees of freedom associated with the other brane stacks, but also the entanglement with high energy modes associated with the given brane stack.

\bigskip

The plan of this paper is as follows. We discuss the field theory definitions of field space entanglement entropy in section \ref{two} and global symmetry entanglement entropy in section \ref{three}. We derive effective descriptions for Coulomb branch and near extremal AdS black holes in sections \ref{d3} and \ref{sec:five}. We explore geometric measures of entanglement and their interpretations in \ref{sec:six} and we conclude in section \ref{sec:conc}. Technical results required for Kaluza-Klein holography for M2-branes and M5-branes are contained in the appendices \ref{appa} and \ref{appb}.

\section{Field space entanglement entropy} \label{two}

 Consider a field theory which may be viewed as two weakly interacting conformal field theories such that the total action is 
\be
I = \int d^{d}x \sqrt{-g} \left ( {\cal L}_{CFT1} + {\cal L}_{CFT2} + g {\cal L}_{int} \right ).  \label{int2}
\ee
In the limit of $g=0$ the Hilbert space factorizes into the direct product of two CFT Hilbert spaces; $g$ controls the interactions between the fields in the two CFTs and should be viewed as being small in all that follows. Note that the UV behaviour of the full theory is controlled by the interactions; since we are interested in holographic realizations, we will mostly consider theories which are UV conformal.

In the interacting theory we can define an entanglement entropy between the degrees of freedom contained in each conformal field theory by tracing out the total density matrix $\rho$ over the degrees of freedom of either:
\be
S^{F} = - {\rm Tr} (\rho_1 \log \rho_1);  \qquad \rho_1 = {\rm Tr}_{CFT2} [\rho]. 
\ee
Operationally we implement the trace by integrating out the fields of the second CFT. 
The entanglement entropy thus defined is clearly qualitatively different from the more familiar entanglement entropy between two different spatial regions of a field theory. For the latter, in any local field theory, only degrees of freedom close to the separating surface are entangled and therefore the leading UV divergence of the entanglement entropy scales with the area of this surface. 

If one defines an entanglement entropy by integrating out degrees of freedom, the remaining degrees of freedom are entangled with those which were integrated out everywhere in the space and therefore one expects the leading UV behaviour of this entanglement entropy to scale with the volume. In this paper, we will denote the entanglement entropy obtained by integrating out degrees of freedom as the field space entanglement entropy, $S^F$, to distinguish it from the usual entanglement entropy.

The field space entanglement entropy has been analysed in a number of condensed matter papers \cite{Furukawa:2010nd,Xu:2011gn,Chen:2013kba,Lundgren:2013gba,Lundgren:2014vea} and was recently studied in simple free field models by \cite{Mollabashi:2014qfa}. Examples of such models include scalar fields interacting by off-diagonal mass terms 
\be
I = - \frac{1}{2} \int d^{d} x \sqrt{-g}  \left ( (\partial \phi^1)^2 + (\partial \phi^2)^2 + m^2 (\phi^1 \cos \alpha - \phi^2 \sin \alpha)^2 \right ) \label{toy1}
\ee
or by off-diagonal derivative interactions
\be
I =  - \frac{1}{2}  \int d^{d} x \sqrt{-g}\left ( (\partial \phi^1)^2 + (\partial \phi^2)^2 + \mu (\partial \phi^1) (\partial \phi^2) \right ). \label{toy2}
\ee 
Entanglement entropy in both models can be computed explicitly by integrating out the field $\phi^2$. Following \cite{Callan:1994py,Holzhey:1994we} one computes the entanglement entropy as a limit of Renyi entropies $S^{F(n)}$, defined as
\be
S^{F(n)} = \frac{1}{(1-n)} {\ln} {\rm Tr} ( \rho^n_{\phi_1}), 
\ee
with
\be
S^{F} = - \left ( \frac{\partial}{\partial n}  {\ln} {\rm Tr} ( \rho^n_{\phi_1} ) \right )_{n=1}. 
\ee
For \eqref{toy2} the entanglement entropy in the ground state behaves as
\be
S^{F} = s(\mu) \left ( c_{d-1} \frac{V_{d-1}}{\epsilon^{d-1}} + \cdots + c_0 \right )
\ee
where $\epsilon$ is the UV cutoff length, $s(\mu)$ is a function of the dimensionless coupling between the fields which vanishes when $\mu =0$, $V_{d-1}$ is the volume of the spatial sections, $c_{l}$ are constants with $c_0$ potentially capturing universal behaviour. For \eqref{toy1} the leading UV divergences are given by
\bea
S^F & \sim & m^4 \sin^2 ( 2 \alpha) V_{d-1} (\ln \Lambda)^2 \qquad d = 5;  \\
S^{F} & \sim & m^4 \sin^2(2 \alpha) V_{d-1} \Lambda^{d-5} (\ln \Lambda) \qquad d \ge 6. \nn
\eea
Note that for $d \le 4$ the entanglement entropy is UV finite. The entanglement entropy vanishes for $m^2 = 0$ (all mass terms vanish) and for $ \alpha = n \pi$ (mass terms diagonal). It scales with the spatial volume, as anticipated, and the powers of $\Lambda$ are consistent with dimensional analysis. 

\bigskip

In general when $g \neq 0$ in \eqref{int2} the entanglement of the ground state follows from the fact that the state cannot be written as a product state
in the Hilbert space which is the product of the Hilbert spaces associated with each decoupled field theory. One can demonstrate this easily in the free field examples: the model with off-diagonal mass terms \eqref{toy1} is solved by diagonalising the mass term, i.e. 
\be
I =  \frac{1}{2} \int d^{d} x \sqrt{-g}  \left ( (\partial \bar{\phi}^1)^2 + (\partial \bar{\phi}^2)^2 + m^2 (\bar{\phi}^1)^2 \right ), \label{toy1-re}
\ee 
with $\bar{\phi}_1 = \phi_1 \cos \alpha - \phi_2 \sin \alpha$ and $\bar{\phi}_2 =  \phi_1 \sin \alpha + \phi_2 \cos \alpha$. The Hilbert space of the theory is therefore diagonal with respect to the fields $\bar{\phi}_1$ and $\bar{\phi}_2$:
\be
{\cal H} = {\cal H}_{\bar{\phi}_1} \otimes {\cal H}_{\bar{\phi}_2}
\ee
but it is not diagonal with respect to the original fields $\phi_1$ and $\phi_2$. 

Note that when $g=0$ in \eqref{int2} the field space entanglement entropy vanishes provided that the theory is in a pure state which is not entangled between the two CFTs. Field space entanglement entropy would not be zero for non-interacting CFTs entangled in thermofield double states or more generally whenever the field theory is in a mixed state.

\subsection{Holographic realization}

It is not a priori clear whether a system of the type \eqref{int2} can be realized holographically. The degrees of freedom associated with the two CFTs are interacting directly. From the field theory perspective it makes sense to view the complete quantum field theory in terms of two interacting subsystems only if the interactions between the two sets of degrees are freedom are very weak relative to their self-interactions, i.e. $g$  is small. 

To obtain a geometric description of the entanglement entropy in holography we will need to impose a similar but inequivalent condition: we require that the two sets of degrees of freedom can be thought of as localised in different regions of the bulk spacetime. While the two regions of the spacetime are in causal contact, we will consider situations in which one can decouple one region to integrate out degrees of freedom. 
In particular we will consider examples for which one can view the bulk spacetime as having an interior decoupling region which is an asymptotically $AdS_{d+1}$ spacetime. The effect of tracing out degrees of freedom is equivalent to specifying particular boundary conditions for this asymptotically $AdS_{d+1}$ spacetime - we will show that these boundary conditions correspond to irrelevant deformations of the effective $d$-dimensional CFT dual to the $AdS_{d+1}$ region.

There has been considerable discussion in earlier literature concerning how interacting conformal field theories should be modelled holographically. One proposal advocates describing a geometry for two conformal field theories interacting via massless modes via Anti-de Sitter spacetimes joined at their conformal boundaries \cite{Aharony:2006hz,Kiritsis:2006hy}. It has also been suggested that conformal field theories interacting via massive modes should be described by two asymptotically AdS bulk geometries glued along a finite size surface in the interior, with their asymptotic conformal boundaries identified \cite{Mollabashi:2014qfa}. Note that the case of interacting CFTs is qualitatively different to the case of non-interacting CFTs entangled in a thermofield double state, described first in \cite{Maldacena:2001kr}.

\bigskip

In this paper we will focus on well-understood classes of holographic geometries which are known to admit field theory interpretations as systems of the type \eqref{int2}, and in which the resulting UV theory is conformal. Our main examples are Coulomb branch geometries, which are engineered using branes that share world volume directions but are distributed over the transverse space. We will primarily consider the case where there are two stacks of branes, $N_1$ and $N_2$ in each stack, separated by a spatial distance, although our analysis could straightforwardly be generalised to additional stacks of branes. 

The conformal symmetry is spontaneously broken and, from the field theory perspective, the low energy degrees of freedom are the massless modes associated with the stack of $N_1$ branes and the massless modes associated with the stack of $N_2$ branes; integrating out the massive modes associated with strings stretched between the branes gives rise to interactions between the two sets of massless modes. The resulting low energy theory is therefore indeed of the form \eqref{int2}.

In Coulomb branch solutions the decoupled geometries are asymptotically $AdS$, but the throat bifurcates into internal throats associated with the locations of the brane stacks, see Figure 1. Deep inside each throat, the geometry is again $AdS$, with a smaller curvature radius, and the low energy field theory description is a CFT.  However, this CFT is not completely decoupled: from the low energy perspective, the field theory is deformed by irrelevant operators, corresponding to integrating out the massive string modes connecting the stacks of branes and in addition integrating out the modes localised at each other brane stack. 

\begin{figure}[h!]
\begin{center}
\includegraphics[scale=2]{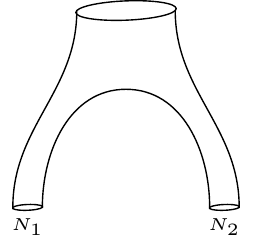}
 \caption{Two throat regions, associated with stacks of $N_1$ and $N_2$ branes.}
\end{center}
\end{figure}

Now let us turn to entanglement entropy. The Ryu-Takayanagi formula describes the entanglement entropy between two spatial regions A and B as the area of the minimal surface in the bulk homologous to the boundary separating the two regions. One way to understand the origin of this formula is by viewing the minimal surface as separating the bulk into two regions, one which can be constructed using only the information in region A and the other being its complement. The leading contributions to the entanglement entropy arise from local interactions at the boundary between the two regions and therefore it is natural that the entanglement entropy is related to the extension of this surface into the bulk. 
In the situation being considered here, the bulk can be divided into two regions, the inner throat region and its complement. It is natural to postulate that the inner throat region should be reconstructable from a reduced density matrix, obtained by integrating out fields, while the complement cannot be constructed from this reduced density matrix. The geometry of this inner throat region should thence be related to field space entanglement entropy. We will propose a general description of the bulk geometry of such a system in section \ref{sec:six}, building on examples given in sections \ref{d3} and \ref{sec:five}, and we will discuss holographic measures of entanglement  in section \ref{sec:six}.

\section{Global symmetry entanglement entropy} \label{three}

\subsection{Generalized holographic entanglement entropy}

In \cite{Mollabashi:2014qfa}  a new holographic functional probing the dependence of the entanglement entropy on the compact part of the geometry was proposed; this was denoted the generalized holographic entanglement entropy. Consider a static spacetime which is asymptotically $AdS_{d+1} \times S^{p}$. The proposed functional is the volume of a minimal codimension two (spatial) hypersurface:
\be
S^G = \frac{1}{4 G_{N}} \int_{\Sigma} d^{d+p-1} x \sqrt{\gamma}, \label{newf}
\ee
where $G_{N}$ is the Newton constant (in $(d+p+1)$ dimensions). The hypersurface $\Sigma$ is chosen to be a minimal hypersurface of constant time with
 boundary conditions such that it completely fills the $(d-1)$-dimensional spatial part of the conformal boundary of $AdS_{d+1}$ and wraps a $(p-1)$-dimensional submanifold of $S^p$. Parameterizing the sphere $S^p$ as
\be
d\Omega_p^2 = d \theta^2 + \sin^2 \theta d \Omega_{p-1}^2,
\ee
then one can for example choose such a submanifold to be an $S^{p-1}$ at a fixed angle $\theta_o$ at the conformal boundary.  The hypersurface $\Sigma$ therefore partitions the spacetime along the internal space at the conformal boundary. In general, as for the usual entanglement entropy, one would expect that a homology constraint is also required but the minimal hypersurface will not be ambiguous in any of the examples discussed here. 

We will use the notation of $S^{G}$ for the quantity \eqref{newf}, again to distinguish it from the standard holographic entanglement entropy obtained from partitioning the spatial background of the dual field theory. Note that the same functional with different boundary conditions has been proposed to evaluate the standard entanglement entropy in top-down models \cite{Hubeny:2007xt}: for the latter one imposes boundary conditions such that the minimal hypersurface completely fills the sphere $S^p$ and partitions the $(d-1)$-dimensional spatial part of the conformal boundary of $AdS_{d+1}$\footnote{It is an interesting open issue to understand the relation of this definition to the usual Ryu-Takayanagi definition in terms of the $(d+1)$-dimensional Einstein metric, since the latter is in general not simply related to the higher-dimensional metric.}.

In \cite{Mollabashi:2014qfa}  it was proposed that the generalised holographic entanglement entropy $S^G$ should be the dual description of the field space entanglement entropy $S^F$. 
There are however various issues with such an interpretation of \eqref{newf}. The most fundamental problem is that the definition of field space entanglement entropy does not assume that the field theory has global symmetry. Any holographic description of this quantity should therefore not need to make explicit reference to the compact part of the geometry, whose existence relies on global symmetry. 
\bigskip

Consider the evaluation of the holographic functional \eqref{newf} in the $AdS_5 \times S^5$ background. It is straightforward to show that the surface $\theta = \pi/2$ is a solution of the minimal surface equations; this follows on symmetry grounds. (For other values of $\theta_o$ the minimal hypersurface is non-trivial and pinches at a finite value of the radius.) Now the area of the spatial surface $\theta = \pi/2$ gives
\be
S^G = \frac{1}{4 G_{N}} \int d^3x \int d \Omega_4 \int R^2 r^2 dr = \frac{2 \pi^2 R^2 V_3}{3 G_N} \int r^2 dr = \frac{2 \pi^2 R^2 r_c^3 V_3}{9 G_N},
\ee
where $V_3$ is the regulated volume of the spatial section and the volume of a 4-sphere is $8 \pi^2/3$. The radial integral is clearly UV divergent and $r_c$ is the radial cutoff. 
Using the standard holographic relation $R^8/G_N  = 2 N^2/\pi^4$ and setting $r_c = \Lambda R^2$, where $\Lambda$ is the cutoff, one obtains
\be
S^G = \frac{4 N^2 V_3}{9 \pi^2} \Lambda^3. \label{ads5}
\ee
This quantity scales with the volume of the spatial section of the field theory, and is UV divergent, just as for the field space entanglement entropy. 
The holographic functional therefore gives a non-zero value for a conformal field theory with R symmetry in its ground state, with the answer depending on how the compact part of the geometry is partitioned; different values are obtained according to the minimal surface determined by the boundary condition $\theta_o$ \cite{Mollabashi:2014qfa}. This is in contradiction to the field space entanglement entropy of section \ref{two}, which is zero in the ground state of any CFT. 

In \cite{Mollabashi:2014qfa} an interpretation of the above result in terms of D3-branes was suggested: consider a spherically symmetric distribution of N D3-branes in $R^6$ such that a plane dividing the $R^6$ has $N/2$ branes on each side. It was then proposed that the generalised holographic entanglement entropy corresponds to the field space entanglement entropy between the two sets of $N/2$ branes. 

However, this interpretation assumes that the conformal ground state of the theory can be viewed as a limiting case of a Coulomb branch solution and hence that the $SO(6)$ symmetry is only approximate at infinite $N$. Yet the ground state of $N = 4$ SYM is $SO(6)$ invariant at any value of $N$ so we cannot view it as being a discrete spherically symmetric distribution of branes.  Other issues concerning the relation of generalized holographic entanglement entropy to field space entanglement entropy were discussed in \cite{Karch:2014pma}. 

\subsection{Field theoretic definition}

In holography the compact part of the bulk geometry (usually an $n$-dimensional sphere, $S^n$) is necessary to capture the global symmetry group of the dual field theory. Therefore the proposed generalised holographic entropy functional should only be applicable to field theories which have global  symmetry. 

Let us briefly review relevant features of field theories dual to holographic geometries. The best understood holographic correspondences include the non-conformal branes, dual to maximally supersymmetric $SU(N)$ Yang-Mills, for which the bosonic terms in the Lagrangian are
\be \label{SYM}
S = \int d^d x \sqrt{-g} {\rm Tr} \left ( - \frac{1}{4 g_d^2} F_{ij} F^{ij} - \frac{1}{2} D_i \phi^a D^i \phi^a + \frac{g_d^2}{4} \sum_{a,b} \left [ \phi^a , \phi^b \right ]^2 \right )
\ee
where $i = 0, \cdots, (d-1)$, $D_i = \partial_i - i A_i$ and there are $n = (10-d)$ scalars, so the index $a$ runs from $1$ to $(10 -d)$. The trace is over $SU(N)$ and the gauge coupling $g_d^2$ is dimensional for $d \neq 4$. The theory has a global $SO(10-d)$ symmetry, under which the gauge fields are singlets and the scalars transform in the fundamental representation; the fermions transform in spinor representations. The holographic dual geometries for the trivial vacua of these theories are conformal to $AdS_{d+1} \times S^{10-d}$; the isometry groups of the bulk spheres are associated with the global symmetry groups of the field theories. 

As a simpler prototype model with global symmetry, we can consider a free field theory with $U(1)$ invariant mass terms such that 
\bea
S_{\alpha} &=& - \frac{1}{2} \int d^{d} x \sqrt{-g} \left ( (\partial \phi^1)^2 + (\partial \phi^2)^2 \right ) \label{toy1a} \\
&& \qquad  -  \frac{1}{2} \int d^{d} x \sqrt{-g} {m}^2 \left  (  (\phi^1)^2 + (\phi^2)^2  \right ). \nn 
\eea
This example can straightforwardly be generalised to a model with $n$ scalar fields $\phi^a$ of equal mass such that $\phi^a$ transforms in the fundamental representation of $SO(n)$.

Now let us turn to the duality between $AdS_3 \times S^3 \times T^4$ and the D1-D5 CFT. 
The Higgs branch of the D1-D5 theory flows in the infrared to an ${\cal N} = (4,4)$ SCFT on $T^4 \times (\td{T}^4)^{N_1 N_5}/S(N_1 N_5)$, where $N_1$ and $N_5$ are the numbers of D1 and D5 branes, respectively, and $S(N_1 N_5)$ denotes the permutation group.  The SCFT on $T^4$ is free but the symmetric product contains interesting dynamics. The SCFT on the orbifold can be described by the Lagrangian
\be
S =  \frac{1}{2} \int d^2 z \left [ \partial x^a_A \bar{\partial} x^a_A  + \psi^a_A \tilde{\partial} \psi^a_A  + \tilde{\psi}^a_A \partial \tilde{\psi}^a_A \right ] \label{D1D5cft}
\ee
Here we switch to Euclidean signature and $a$ runs over the $\td{T}^4$ coordinates while $A= 1, \cdots N_1 N_5$ labels the copies of the torus. The symmetric group acts by permuting the copy indices and introduces twisted sectors. The theory has an $SO(4) = SU(2)_L \times SU(2)_R$ global R symmetry under which the  bosons transform as $(2,2)$ while the fermions transform as $(2,1)$ and $(1,2)$; this symmetry corresponds to the isometry group of $S^3$ in the holographic dual $AdS_3 \times S^3 \times T^4$. The theory also has a local $SU(2) \times \tilde{SU(2)}$ R parity symmetry. 

A simpler prototype would be the bosonic theory with $N_1 N_5 = 2$:
\be
S = \frac{1}{2} \int d^2 z \left [ \partial x^a_A \bar{\partial} x^a_A \right ],
\ee
where now $A = 1,2$ and $a$ still runs over the $\td{T}^4$ coordinates. This theory still admits the global $SO(4)$ symmetry associated with rotations of the scalars but does not have a local R symmetry. 

\bigskip

In free field theories such as the toy models discussed above one can define a global symmetry entanglement entropy as follows. Let $\rho$ be the density matrix of the theory, which we first take to be a pure quantum state $| \Psi \rangle$ so that $\rho = | \Psi \rangle \langle \Psi |$. One can always define a reduced density matrix by tracing out degrees of freedom. In the free field models we defined the field space entanglement entropy using the reduced density matrix obtained by integrating out one of the scalars. We now define a global symmetry entanglement entropy in the toy model \eqref{toy1a} by constructing the reduced density matrix
\be
\rho_{\lambda} = \int_{{\cal S}_{\lambda}} D \phi_1 D \phi_2 | \Psi \rangle \langle \Psi |
\ee 
and then defining
\be
S^R (\lambda) = - {\rm Tr} (\rho_{\lambda} \ln \rho_{\lambda}).
\ee
Here ${\cal S_\lambda}$ is a wedge of angle $\lambda$ in the $R^2$ in which the real fields $(\phi^1,\phi^2)$ take values. 

The generalisation to a theory with a number $n$ of scalars $\phi_a$ transforming in the fundamental of $SO(n)$ is immediate. Let $\phi_a = \phi \hat{n}_a$ where $\hat{n}_a$ such that $\hat{n}^a \hat{n}^a = 1$ is a normal vector to the unit $S^{n-1}$. Now we can define a reduced density matrix and an entanglement entropy by defining a spherical cap ${\cal S}$, i.e. by choosing a plane which bisects the unit $S^{n-1}$, and integrating out all field configurations on one side of the cap:
\be
\rho_{\cal S} = \int_{\phi_a: \hat{n}_a \in \cal S} D \phi_a | \Psi \rangle \langle \Psi |; \qquad
S_{\cal S} = - {\rm Tr} (\rho_{\cal S} \ln \rho_{\cal S}).
\ee
In principle this definition does not rely on the field theory being non-interacting although in practice it would difficult to implement the integration in an interacting theory, even without the additional complications of gauge freedom implicit in theories such as \eqref{SYM}. 

\subsection{Evaluation for free fields}

As a warm up, we consider two massive scalars in the case of $d=1$, i.e. quantum mechanics. The ground state wave function is then
\be
\psi( \{ \phi \} ) \equiv \langle \{ \phi \} | \Psi \rangle = \frac{m^{1/2}}{{\pi}^{1/2}} {\rm exp} \left [ - \frac{1}{2} m ( \phi_1^2 + \phi_2^2) \right ] 
\ee
and the corresponding density matrix is
\be
\rho (\{ \phi \}, \{ \td{\phi} \}) \equiv \psi(\{ \phi \} ) \psi(\{ \td{\phi} \} )^{\ast} = \frac{m}{{\pi}} {\rm exp} \left [ - \frac{1}{2} m ( \phi_1^2 + \phi_2^2 + \td{\phi}_1^2 + \td{\phi}_2^2) \right ] 
\ee
which is clearly normalised to satisfy the condition ${\rm Tr}(\rho) = 1$. The reduced density matrix is obtained by integrating out the wedge. It consists of two parts:
\be
\rho_{\rm wedge}  =   \frac{m}{{\pi}}  \int_{ {\cal S}_{\lambda}}  d\phi_1 d \phi_2 {\rm exp} \left [ - m ( \phi_1^2 + \phi_2^2  ) \right ] = \frac{\lambda}{2 \pi}.
\ee
together with 
\be
\rho_{\rm outside} = \frac{m}{{\pi}} {\rm exp} \left [ - \frac{1}{2} m ( \phi_1^2 + \phi_2^2 + \td{\phi}_1^2 + \td{\phi}_2^2) \right ].
\ee
Thus
\be
\rho_{\lambda} = \rho_{\rm wedge} + \rho_{\rm outside}
\ee
By construction ${\rm Tr} (\rho_\lambda) = 1$.
We now compute the entanglement entropy using
\be
S^R (\lambda) = - \frac{d}{dn} \left ( {\rm ln} {\rm Tr} (\rho_\lambda^n) \right )_{n=1}.
\ee
Since
\be
{\rm Tr} (\rho_{\lambda}^n) = \left ( \frac{\lambda}{2 \pi} \right )^n + \left ( \frac{2 \pi - \lambda}{2 \pi} \right )^n
\ee
we find that 
\be
S^R(\lambda) = - \frac{\lambda}{2 \pi} \log \left ( \frac{\lambda (2 \pi - \lambda)}{4 \pi^2} \right ).
\ee
This vanishes as $\lambda \rightarrow 0$ and gives $S^R = \log (2)$ for $\lambda = \pi$, which corresponds to a partitioning of the field space into two. For $\lambda = 2 \pi ( 1- \delta)$ with $\delta \ll 1$, 
\be
S^R (2 \pi (1- \delta)) \approx - \log (\delta)
\ee
which diverges as $\delta \rightarrow 0$, since in this limit all of the fields are integrated out.  

\bigskip

We now consider the generalisation to two equal mass scalars in $d > 1$. The ground state wave function is 
\be
\psi( \{ \phi \} ) \equiv \langle \{ \phi \} | \Psi \rangle = {\cal N} {\rm exp} \left [ - \frac{1}{2} \int d^{d-1} x d^{d-1} y W(x,y) ( \phi_1(x) \phi_1(y)  + \phi_2(x) \phi_2 (y) ) \right ], 
\ee
where ${\cal N}$ is a normalisation factor and 
\be
W(x,y) = V_{d-1} \sum_{k} (k^2 + m^2)^{1/2} e^{ik (x-y)},
\ee
with $V_{d-1}$ the spatial volume. The normalisation factor is determined by the condition ${\rm Tr}(\rho) = 1$ to be
\be
{\cal N} = {\rm det} ( \pi^{-1} {\rm Re}(W))
\ee
The reduced density matrix is now defined using 
\be
\rho_{\rm wedge} = \int_{\cal S_{\lambda}} D \phi_1 D \phi_2  \langle \{ \phi_1, \phi_2 \} | \Psi \rangle \langle \Psi | \{ {\phi}_1, \phi_2 \} \rangle 
\ee
with ${\cal S}_{\lambda}$ denoting the wedge region and
\be
\rho_{\lambda} = \rho_{\rm wedge} + \rho_{\rm outside} (\phi_1, \tilde{\phi_1}, \phi_2, \tilde{\phi}_2).
\ee
Again the entanglement entropy is easiest to compute for $\lambda=\pi$, corresponding to the partitioning of the field space into two halves, in which case
\be
{\rm Tr} (\rho_{\pi}^n) = 2^{1-n} = 2^{-\epsilon}
\ee
where $n = 1 + \epsilon$. By construction ${\rm Tr} (\rho_\lambda) = 1$. Thus
\be
S^R(\pi) = \log 2,
\ee
which is again UV finite. The entanglement entropy is finite in this case since the degrees of freedom which have been traced out in the reduced density matrix are not interacting with the remaining degrees of freedom. 

\subsection{Interactions}

Consider a more general field theory in which the scalar fields are interacting. Integrating out fields within the spherical cap thus should give rise to an entanglement entropy which scales with spatial volume and is UV divergent, since all remaining degrees of freedom are entangled with those which were integrated out, at all scales. Such behaviour is qualitatively different from that in the free field models of the previous section. The simplest prototype for the behaviour in SYM would therefore be an interacting theory with global symmetry which is exactly solvable and for which the ground state preserves the global symmetry.  

In both of the toy models \eqref{toy1} and \eqref{toy2} there are interactions between the two species of scalar fields but these can be solved by carrying out orthogonal transformations on the fields, so that the resulting theory is free. Both models have a $U(1)$ R symmetry which is explicitly broken by the interactions; in the massive field case the theory is nonetheless UV conformal (free), but in the derivative interaction case the R symmetry is broken at all scales. As discussed earlier the latter is a reasonable prototype for the Coulomb branch of SYM (although the R symmetry is spontaneously rather than explicitly broken in the latter) but neither toy model is an ideal prototype for SYM in the R symmetric vacuum. Nonetheless, these results suggest that for SYM the leading terms in the global symmetry entanglement entropy should be of the form
\be
S_{\cal S} \sim s ({\cal S}) V_{d-1} \Lambda^{d-1}, 
\ee
where $s(\cal S)$ depends on the partition of the symmetry orbit. This would be in qualitative agreement with \eqref{ads5}. 

\subsection{General definition of global symmetry entanglement entropy}

The definition of global symmetry entanglement entropy would be subtle in gauge theories and in theories with other elementary fields such as fermions transforming in non-fundamental representations of the global symmetry group. In a gauge theory the scalar fields would not be gauge invariant and therefore our construction should be replaced by a manifestly gauge invariant procedure. In defining the reduced density matrix, it is unclear how  one should trace out fermionic degrees of freedom. In an interacting theory, it is not clear that the Hilbert space can be expressed as a direct product, as in the free field models. Note however that analogous issues affect the usual entanglement entropy both in gauge theories and in theories with fermions. 

More generally, to make contact with holography, the definition of global symmetry entanglement entropy would better be expressed in terms of gauge invariant operators rather than elementary fields, since the latter do not exist in holographic realisations. In a conformal field theory with global 
symmetry, the most natural basis for gauge invariant operators is R symmetry eigenstates. Global symmetry entanglement entropy is obtained by integrating out part of the orbit of the R symmetry, and therefore its definition requires states which are localised along orbits of the R symmetry, rather than eigenstates of R symmetry. 

To illustrate this, consider a conformal field theory with a $U(1)$ global symmetry, in which orthonormal states are labelled by their conformal dimensions $\Delta$, their R charge $n$ and additional degeneracy labels $k$, $| \Delta; n; k \rangle$. 
A new orthonormal basis can always be defined as a superposition
\be
| \psi^{\alpha}  \rangle = \sum_{\Delta,n,k} \alpha_{\Delta,n,k} | \Delta; n; k \rangle \label{suer}
\ee
where by construction the expansion coefficients satisfy
\be
\langle \psi^{\beta}  | \psi^{\alpha}  \rangle = \sum_{\Delta,n,k} \beta^{\ast}_{\Delta,n,k} \alpha_{\Delta,n,k} =0. 
\ee
If such a conformal field theory is realised holographically in $AdS_{d+1} \times S^1$, then the spectrum of states is usually expressed in terms of R symmetry eigenstates, i.e. modes with definite angular momenta along the circle. For every operator creating a state $| \Delta; n; k \rangle$ there is a corresponding dual $(d+1)$-dimensional field $\Phi_{\Delta;n;k}$, which in turn is associated with a $(d+2)$-dimensional mode carrying momentum $n$ along the $S^1$. By superposing such modes one can create fields which are localised along the circle; such a procedure would determine the coefficients in \eqref{suer}. Given the basis $| \psi^{\alpha} \rangle$ of states localised along the R symmetry orbit, one could then trace out states localised within part of the orbit, and hence define a global symmetry entanglement entropy. 

This is very similar to the proposal made in \cite{Karch:2014pma}. However, we should note that it would be very hard to compute the coefficients in \eqref{suer} in practice even in cases for which the detailed map between bulk Kaluza-Klein fields and boundary operators is known: the dictionary between spherical harmonics of ten or eleven dimensional supergravity fields and boundary operators is complicated and highly non-linear, see for example \cite{Skenderis:2006uy,Skenderis:2006di,Skenderis:2007yb}.

\section{D3-brane supergravity solutions} \label{d3}

In this section we will explore separated D3-brane stacks, i.e. Coulomb branch solutions of ${\cal N} = 4$ SYM. We will argue that there is an effective five-dimensional description of the throat geometry near a given brane stack, corresponding to the field theoretic description of a CFT deformed by irrelevant operators. Earlier discussions of entanglement in this system can be found in \cite{Mollabashi:2014qfa,Aprile:2014iaa,Karch:2014pma}.

Supergravity solutions for D3-branes on the Coulomb branch can be expressed as 
\bea
ds^2 &=& H(y)^{-1/2} dx \cdot dx + H(y)^{1/2} dy^a dy_a; \\
F_{5} &=& d C_4 + \ast_{R^6} d C_4; \qquad C_4 = H(y)^{-1} dx_4, \nn
\eea
where $H(y)$ is a harmonic function on $R^6$:
\be
H(y) = \sum_{l=1}^{N} \frac{4 \pi \alpha'^2 g_s}{| y - y_l|^4},
\ee
and $y_l$ denote the locations of each D3-brane. Note that the supergravity solutions are non-singular only if the distribution of D3-branes is continuous, on a compact hypersurface of dimension four or less. Solutions involving separating stacks of branes are mildly singular, as one can only remove the singularity at a single stack, see \cite{Klebanov:1999tb} and the discussion below. Nonetheless let us consider the case of two stacks of equal charge\footnote{For computational simplicity in this section we split the branes into two equal charge stacks but the generalisation to stacks of different charge would be straightforward.}, separated by distance $2l$ along the $y^1$ direction. In this case the function $H(y)$ is
\be
H(y) = \frac{2 \pi g_s \alpha'^2 N}{(r^2 + l^2 - 2 r l \cos \theta)^2} + \frac{2 \pi g_s \alpha'^2 N}{(r^2 +l^2 + 2r l \cos \theta)^2} \label{hdef}
\ee
and the geometry preserves an $SO(5)$ subgroup of the $SO(6)$ symmetry group of the $S^5$. Here we have introduced spherical polar coordinates for the $R^6$ such that $y^1 = r \cos \theta$ and 
\be
dy^a dy_a = dr^2 + r^2 (d \theta^2 + \sin^2 \theta d \Omega_4^2)
\ee
with $0 \le \theta < \pi$. 

As argued by  \cite{Klebanov:1999tb}, in the neighbourhood of each stack the geometry is $AdS_5 \times S^5$, with a radius appropriate to $N/2$ branes. To see this, let 
\be
r = l + z, \qquad \theta = \delta
\ee
with $\delta \ll z $. Then 
\be
H \approx \frac{R^4}{2 z^4} + \frac{R^4}{2 (2l + z)^4} \label{nh}
\ee
where we have set $R^4 = 4 \pi g_s \alpha'^2 N$. The first term gives an AdS warp factor with radius $R/2^{1/4}$, appropriate to that of $N/2$ branes. The additional term can be understood in terms of irrelevant deformations of the $SU(N/2)$ CFT \cite{Intriligator:1999ai,Skenderis:2006di}. 

As pointed out in \cite{Klebanov:1999tb}, the metric defined by \eqref{hdef} is singular at the locations of both stacks of branes. Using \eqref{nh}, the metric in the vicinity of the stack at $r = l$ (i.e. assuming $z \ll l$) is
\be
ds^2 = \left (\frac{R^4}{2 z^4} \right  )^{-1/2} dx \cdot dx + \left (\frac{R^4}{2 z^4} \right )^{1/2} (dz^2 + l^2 d \Omega_5^2)
\ee
which is clearly singular as $z \rightarrow 0$, since the warp factor of the $S^5$ diverges. To remove the singularity one shifts the stacks of branes (or, equivalently, redefines coordinates) so that one stack is at $y = 0$ and the other stack is at $y^1 = -2l$. Then 
\be
H(y) = \frac{R^4}{2 r^4} + \frac{R^4}{2 (r^2 + 4 l^2 + 4 r l \cos \theta)^2}. \label{sw}
\ee
The resulting metric is regular as $r \rightarrow 0$ but is still singular at $y^1 = -2 l$; one can only remove the singularity at one stack by coordinate transformations.  Since one cannot eliminate all singularities, we will work with the form of the metric \eqref{hdef}, but we need to bear in mind the naked singularities at $y^{1} = \pm l$. 

\subsection{Generalized holographic entanglement entropy}

In this section we briefly review the evaluation of the generalized holographic entanglement entropy \eqref{newf} for the Coulomb branch solutions, highlighting various subtleties relative to \cite{Mollabashi:2014qfa}. The proposed functional \eqref{newf} was evaluated for the D3-brane stacks discussed above: the surface $\theta = \pi/2$ is a solution of the minimal surface equations (by symmetry). Then 
\be
S^G = \frac{1}{4 G_{N}} \int d^3x \int d \Omega_4 \int \frac{R^2 r^4 dr}{(r^2 + l^2)} = 
\frac{2 \pi^2 R^2 V_3}{3 G_N}  \int \frac{ r^4 dr}{(r^2 + l^2)}. \label{g-one}
\ee
The radial integral was regulated in \cite{Mollabashi:2014qfa} as $0 \le r \le r_c \ll l $, which is the red region in Figure 2. (Note that $\theta = 0$ on the positive $y^1$ axis; $\theta = \pi$ on the negative $y^1$ axis and $\theta = \pi/2$ on the orthogonal axis.) Then
\be
S^G =
\frac{2 \pi^2 R^2 V_3}{3 G_N}  \frac{ r_c^5}{5 l^2}.
\ee

\begin{figure}[h!]
  \centering

\includegraphics[scale=0.8]{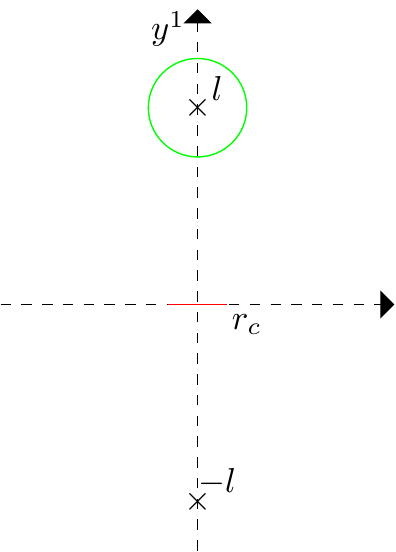}

  \caption{The $(r,\theta)$ plane is shown, with the red region indicating the integration of \cite{Mollabashi:2014qfa} while the green region is the cutoff for our analysis. The locations of the brane stacks are indicated with crosses.}
   
\end{figure}

In the next steps the authors of \cite{Mollabashi:2014qfa} apply the standard holographic relations $R^8/G_N  = 2 N^2/\pi^4$ and $R^2 = 2 \pi \alpha' \sqrt{\lambda}$ together with 
\be
r_c = \Lambda R^2 \qquad g = (\pi \alpha'^2) \frac{ \Lambda^2}{l^2} \label{cutoff}
\ee
where $\Lambda$ is the UV cutoff. The first equality follows from imposing the standard relation between geometric IR cutoff and field theory UV cutoff. The second relation
introduces a dimensionless coupling $g$ which characterises the interactions between the brane stacks: $g$ becomes of order one when the mass scale set by the brane separation is of order the cutoff scale. Using both these relations in the expression \eqref{g-one} we obtain
\be
S^G =
\frac{16 N^2 V_3}{15 \pi^2 }  \lambda g \Lambda^3.
\ee
This expression correctly gives zero as $g \rightarrow 0$ and qualitatively reproduces the behaviour found in CFTs with massless interactions - the derivative coupling case reviewed in section \ref{two}. 

However, the use of the first expression in \eqref{cutoff} is  conceptually flawed: the UV of the dual field theory corresponds to the region $r \gg l$ for all $\theta$: this would be a circle of large radius in Figure 2, enclosing both brane stacks. The red line does not approach the boundary of the spacetime and therefore $r_c$ cannot be thought of as a UV cutoff. Indeed along the red line the warp factor $H$ is approximately constant and the metric is approximately flat, rather than being approximately $AdS_5 \times S^5$! In other words, we cannot use the AdS/CFT relation when $r_c$ is deep within the spacetime and nowhere near the $AdS$ boundary. In Figure 1, the red region corresponds to the surface deep inside the spacetime at which the throat bifurcates.

\subsection{Irrelevant deformations}

We now consider the interpretation of the bulk geometry in terms of interacting CFTs. Associated with each stack of branes we have $SU(N/2)$ CFTs. The interactions between the CFTs are via massive string modes, with the mass scale being
\be
M = \frac{l}{\pi \alpha'}.
\ee
In the CFT language these massive string modes correspond to irrelevant bifundamental operators in the product of the two CFTs. The strength of the interactions can be characterised by the dimensionless coupling $g = \Lambda^2/M^2$ introduced above. It would only make sense to consider the system as well-described by two weakly interacting CFTs if $g$ is small, which in turn requires that $M$ is large compared to the cutoff scale. It is claimed in \cite{Mollabashi:2014qfa} that the  limit $l \gg r_c$, i.e. the region indicated by red in Figure 2, that we can view the system as two weakly interacting CFTs. However, geometrically it is hard to justify this interpretation as this region (in which the throat bifurcates) is not decoupled and the metric is approximately flat. 

Let us now explore the system further by zooming in on the vicinity of one brane stack. It is convenient to switch to the metric with the defining function being \eqref{sw}. Then we take the limit of $r \ll l$.
This is not the same region as discussed previously, but rather the region enclosed within the green circle of Figure 2\footnote{Equivalently this region is that inside one of the throats of Figure 1.}. Geometrically, the throat region bifurcates into two narrower throats which each approach one stack of branes. 

Expanding the warp factor for $1 \ll r \ll l$ we obtain
\be
H = \frac{R^4}{2 r^4} + \frac{R^4}{32 l^4} \left ( 1 + \frac{r}{l} \cos \theta + \frac{r^2}{4 l^2} \right )^{-2} \approx \frac{R^4}{2r^4} \left ( 1 + \frac{\td{g}^4}{16} + {\cal O} (\td{g}^5)  \right )
\ee
where we define $\td{g} = r/l \ll1$. 
The warp factor produced by the stack of branes at $r=0$ is that for a conformal field theory, with gauge group $SU(N/2)$, as expected. The warp factor produced by the second set of branes corrects $H$ by terms which are small within the inner throat region. 

Substituting into the metric we find
\be
ds^2 = \frac{ \sqrt{2} r^2}{R^2} (1 - \frac{\td{g}^4}{32} + \cdots ) \left (dx \cdot dx \right ) +  \frac{ R^2}{ \sqrt{2} r^2} (1 + \frac{\td{g}^4}{32} + \cdots) \left (dr^2 + r^2 d \Omega_5^2 \right ). \label{met2}
\ee
Using the method of Kaluza-Klein holography (see the final section of \cite{Skenderis:2006di}), we can read off the correspondence between the expansion of $H$ in terms of scalar spherical harmonics $Y_{k}^{I}$ on the $S^5$ and irrelevant deformations of the $SU(N/2)$ CFT associated with the stack of $N/2$ branes. Following formula (6.1) from that paper, we write the warp factor $H$ as 
\be
H = \sum_{k,I} \left ( l_{kI} r^{k} + \frac{h_{kI}}{r^{k+4}} Y_{k}^{I} (\theta) \right ). 
\ee
Here $(l_{kI},h_{kI})$ are expansion coefficients. 
Expressing
\be
H = \frac{R^4}{2 r^4} + \delta H
\ee
we see that our $\delta H$ is expanded in positive powers, i.e. only $l_{kI}$ are non-zero. The scalar $t_{k}^I$ fields are given by
\be
t_{k}^{I} = \frac{ l_{kI}}{4 (k+4)} r^{k+4}.
\ee
For $r \ll \l$, the leading correction term is that for which $k=0$. The $t_{0}$ field corresponds to the R singlet dimension eight operator ${\rm Tr} (F^4)$ and the coefficient $l_{0} \sim \frac{1}{l^4}$ describes a deformation of the CFT by this operator: 
\be
\delta I \propto \frac{1}{l^4} \int d^4 x \sqrt{-g}  {\rm Tr} (F^4).
\ee 
This deformation was first dscussed by Intriligator in \cite{Intriligator:1999ai}, in the context of understanding how the decoupled $AdS_5 \times S^5$ part of the D3-brane geometry can be extended to the asymptotically flat geometry. By construction the irrelevant deformation has a coefficient which is small for energy scales much smaller than $l$. 

The next correction term is that for which $k=1$. The scalar $t_{1}^{I}$ fields correspond to a dimension nine operator transforming in the ${\bf 6}$ of $SO(6)$. The only active scalar is a singlet under an $SO(5) \in SO(6)$ and describes a deformation of the CFT by the dimension nine operator breaking the global symmetry to $SO(5)$, i.e.
\be
\delta I  \propto \frac{1}{l^5} \int d^{4} x \sqrt{-g}  {\rm Tr} (F^4 \phi^1),
\ee
where we label the six scalar fields of $N=4$ SYM as $\phi^a$, as in \eqref{SYM}. 
To go to arbitrarily higher order in the expansion in terms of powers of $1/l$, we would need non-linear terms in the Kaluza-Klein holography dictionary \cite{Skenderis:2006di}. However, it is clear that the general structure is that the $SU(N/2)$ CFT is deformed by irrelevant $SO(5)$ singlet operators with deformation parameters proportional to $l^{4 - \Delta}$. 


Now let us evaluate the generalized holographic entropy functional \eqref{newf} in the metric \eqref{met2}, along a slice of $\theta = \pi/2$. This gives
\be
S^G = \frac{8 \pi^2 V_3 R^2}{3 \sqrt{2} G_{N}} \int_{0}^{r_{\Lambda}} dr r^2 \left (1 + \frac{r^4}{16 l^4} \right )
\ee
The first term reproduces the  form of the expression found in  \cite{Mollabashi:2014qfa} and reviewed above in \eqref{ads5} for pure $AdS_5$:
\be
S^{G(1)} = \frac{N^2}{9 \pi^2} \frac{V_3}{\epsilon^3}
\ee
where we take the cutoff $r_{\Lambda} = \frac{R^2}{\sqrt{2} \epsilon}$ corresponding to $N^2 \rightarrow \frac{N^2}{4}$. Working out the second term in the integral we obtain
\be
S^{G(2)} =  \frac{3 \td{g}^4}{7} S^{G(1)} 
\ee
Clearly, while $S^G$ is sensitive to the irrelevant deformations, and thus to integrating out the other degrees of freedom, it is non-zero even when there is no second stack of branes and the irrelevant deformation vanishes, i.e. $\td{g} = 0$.  This is in line with what we found for the global symmetry entanglement entropy defined in section \ref{three}, again supporting the identification of the generalized holographic entropy with this quantity.  

\subsection{Effective description via dimensional reduction}

In the previous section we argued that the effective geometry at the top of the inner throat associated with a brane stack is anti-de Sitter, plus certain corrections which can be viewed as small provided that our radial cutoff is small compared to the scale set by the irrelevant deformations. 

In holography it is more straightforward to work with an asymptotically AdS geometry than with the uplifted higher-dimensional geometry which is asymptotic to the product of AdS with a compact space. However, a generic higher-dimensional solution cannot be expressed as the uplift of a lower-dimensional solution of a consistently truncated theory. Whenever the geometry is AdS cross a sphere plus small corrections, as in the analysis above, the techniques of Kaluza-Klein holography \cite{Skenderis:2006uy,Skenderis:2006di,Skenderis:2007yb} can however be exploited to construct a lower-dimensional effective description. 
 
In the case at hand we can express the solution in five-dimensional language working perturbatively in the parameter $\tilde{g}$, which is small at the boundary of the inner throat region (i.e. the green regions of Figures 2 and 3).

\begin{figure}[h!]
\begin{center}
\includegraphics[scale=2]{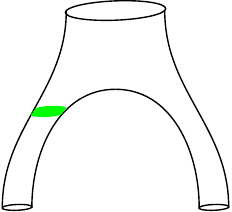}
 \caption{The shaded green region of the inner throat can be described in five dimensional language by small deformations of $AdS$, characterizing the irrelevant deformations of the dual CFT.}
\end{center}
\end{figure}
 
The leading terms in five dimensions are captured by the following action, see \cite{Skenderis:2006uy,Skenderis:2006di,Skenderis:2007yb}:
\be
I = \frac{N^2}{32 \pi^2} \int d^{5}x \sqrt{-g} \left  ({\cal R}  + 12 - \frac{1}{2}  ( (\partial T_0)^2 + 32 T_0^2) + \cdots \right )
\ee
with the solution (to quadratic order in $1/l^4$) of interest being
\bea
ds^2 &=& g(\bar{r}) \frac{d \bar{r}^2}{\bar{r}^2 } + \bar{r}^2 f(\bar{r})  d x \cdot dx;  \label{red-metric} \\
T_0 &=& c \frac{\bar{r}^4}{l^4}, \nn 
\eea
where $c$ denotes a computable numerical constant and the AdS radial coordinate has been rescaled to $\bar{r}$. 
Here the metric functions $g(\bar{r})$ and $f(\bar{r})$ depend on the gauge choice, with only the combination
\be
g(\bar{r}) - \bar{r} \partial_{\bar{r}} f(\bar{r}) = 1 - \frac{2}{3} c^2 \frac{\bar{r}^8}{l^8}
\ee
being determined by the Einstein equations. In the gauge choice $g(\bar{r}) = 1$ this expression implies that
\be
f(\bar{r}) =  1 - \frac{c^2 \bar{r}^8}{12 l^8}.
\ee
Note that the Einstein frame metric in five dimensions is related to the higher dimensional metric by a Weyl rescaling; the lower-dimensional metric has to be AdS to linear order, as the backreaction of the scalar field $T_0$ is of order $\td{g}^8$. The scalar field $T_0$ is a rescaling of the field $t_0$; its mass is appropriate for a dual operator of dimension eight. The terms in ellipses denote additional fields dual to the higher dimension irrelevant operators which break the $SO(6)$ R symmetry to $SO(5)$. For example, the ten-dimensional fields $t^I_1$ reduce to (rescaled) scalar fields $T_1^I$ of mass $45$ in AdS units. 

\section{Other holographic systems} \label{sec:five}

The D3-brane solutions discussed in the previous section describe the flow from  ${\cal N} = 4$ $SU(N)$ SYM to the infrared. The inner throat region in the interior geometry associated with a brane stack was argued to describe an IR conformal field theory, deformed by irrelevant operators. The corresponding effective geometric description in five dimensions is Einstein gravity coupled to massive scalar fields. Such a situation occurs rather generically in holography and in this section we will consider other examples. 

\subsection{M-branes}

The M2-brane and M5-brane geometries for Coulomb branch solutions can be expressed as
\be
ds^2 = H(y)^{-\alpha} (dx \cdot dx)_d + H(y)^{1 - \alpha} (dy \cdot dy)_D
\ee
where $H(y)$ is a harmonic function on $R^D$. Here $\alpha = (D-4)/(D-2)$ and $D = (11- d)$ for the M-branes of worlvolume dimension $d$. For brane stacks separated along the $y^1$ direction by distance $l$ the harmonic function takes the form
\be
H(y) = \frac{Q_1}{r^{D-2}} + \frac{Q_2}{ (r^2 + l^2 + 2 r l \cos \theta)^{\frac{D-2}{2}}}, \label{harm}
\ee
where we place one of the brane stacks at $y^a=0$ and again choose $y^1 = r \cos \theta$ etc. For $r \ll l$
\be
H(y) = \frac{Q_1}{r^{D-2}} + \frac{Q_2}{l^{D-2}} + \cdots
\ee
and hence the geometry is
\bea
ds^2 &=& \frac{\lambda^2}{z^2} \left ( dz^2 + dx \cdot dx_d \right ) + Q_1^{1- \alpha} (1 +  \tilde{g}^d) d \Omega_{D-1}^2 + \cdots \label{mgeom} \\
\lambda &=& \frac{2}{(D-4)} Q_1^{\frac{1}{2} (1 - \alpha)} \nn \\
\tilde{g}^{d} &=& Q_2 \frac{ (1 - \alpha)}{Q_1} \left ( \frac{r}{l} \right )^{D-2} \equiv \left ( \frac{\tilde{l}}{z} \right )^{d} \nn 
\eea
where $\tilde{g}$ is considered to be smaller than one so that the other terms, denoted by ellipses, are subleading. 

We can now use the results on Kaluza-Klein holography for $AdS_4 \times S^7$ and $AdS_7 \times S^4$ contained in appendices \ref{appa} and \ref{appb} to interpret the geometry \eqref{mgeom} in terms of irrelevant deformations
of the dual field theories. Using appendix \ref{appa} for M2-branes, the metric \eqref{mgeom} can be expressed as 
\be
g_{MN} = g^o_{MN} + h_{MN}
\ee
where $g^o_{MN}$ is the $AdS_4 \times S^7$ background and hence from \eqref{mgeom}
\be
h_{MN} dx^{M} dx^{N} = 4 \lambda^2 \tilde{g}^3 d \Omega_7^2 + \cdots
\ee
Using \eqref{appa7} and \eqref{appa15}, we can interpret this change in the metric as a source for a dimension six operator ${\cal O}_{\pi^0}$, i.e. the field theory deformation corresponding to the throat geometry is
\be
I \rightarrow I_{Q_1} + \tilde{l}^3 \int d^3 x {\cal O}_{\pi^0},
\ee
where $I_{Q_1}$ denotes the CFT dual to $Q_1$ M2-branes. 

Following the same arguments as in the previous section, the effective description of the top of the inner throat region in four-dimensional language is thus gravity coupled to the scalar field dual to ${\cal O}_{\pi^0}$, i.e.
\be
I = \frac{1}{16 \pi G_4} \int d^4x \sqrt{-g} \left ( { \cal R} + 3 - \frac{1}{2} ( (\partial \pi^0)^2 + 18 (\pi^0)^2 )  + \cdots \right ),
\ee
where the ellipses correspond to additional Kaluza-Klein modes associated with the subleading terms in the metric \eqref{mgeom}. 

Using appendix \ref{appb} for M5-branes, the metric \eqref{mgeom} can again be expressed as $g_{MN} = g^o_{MN} + h_{MN}$ with $g^o$ the background $AdS_7 \times S^4$ metric and 
\be
h_{MN} dx^M dx^N = \frac{1}{4} \lambda^2 \tilde{g}^6 d \Omega_4^2 + \cdots
\ee
From \eqref{appb5} and \eqref{appb11}, we can interpret this change in the metric as a source for a dimension twelve operator ${\cal O}_{\pi^0}$, i.e. the field theory deformation corresponding to the throat geometry is
\be
I \rightarrow I_{Q_1} + \tilde{l}^6 \int d^6 x {\cal O}_{\pi^0},
\ee
where $I_{Q_1}$ is the action for the CFT dual to $Q_1$ M5-branes. 

The effective description of the throat region in seven-dimensional language is thus gravity coupled to the scalar field dual to ${\cal O}_{\pi^0}$, i.e.
\be
I = \frac{1}{16 \pi G_4} \int d^7x \sqrt{-g} \left ( { \cal R} + 6 - \frac{1}{2} ( (\partial \pi^0)^2 + 72 (\pi^0)^2 )   + \cdots \right ),
\ee
where the ellipses again correspond to additional Kaluza-Klein modes associated with the subleading terms in the metric \eqref{mgeom}. 

\subsection{D1-D5 system}

Extremal D1-D5 geometries can similarly be expressed as solutions to six dimensional supergravity, see \cite{Kanitscheider:2006zf}, with the metric being:
\be
ds^2 = H_1(y)^{-1/2} H_{5}(y)^{-1/2} (dx \cdot dx)_2 + H_1(y)^{1/2} H_5(y)^{1/2} (dy \cdot dy)_4,
\ee
where $H_1(y)$ and $H_5(y)$ are both harmonic functions on $R^4$.  For separated D1-D5 brane stacks, in which the D1 branes and D5 branes remain coincident, each harmonic function takes the form of \eqref{harm}: 
\be
H_1(y) = \frac{Q_1}{r^2} + \frac{Q'_1}{(r^2 + l^2 + 2 r l \cos \theta)}; \qquad
H_5(y) = \frac{{Q}_5}{r^2} + \frac{ {Q}'_5}{(r^2 + l^2 + 2 r l \cos \theta)},
\ee
and thus for $r \ll l$
\be
H_1(y) = \frac{Q_1}{r^2} + \frac{Q'_1}{l^2} + \cdots; \qquad
H_5(y) = \frac{Q_5}{r^2} + \frac{Q'_5}{l^2} + \cdots \label{harm2}
\ee
so the metric becomes
\bea
ds^2 &=& \frac{r^2}{\sqrt{Q_1 Q_5}}( dx \cdot dx) + \frac{\sqrt{Q_1 Q_5} dr^2}{r^2} + \sqrt{Q_1 Q_5} d \Omega_2^2 
\left (1 + \tilde{g}^2 \right ) + \cdots \\
\tilde{g}^2 &=& \frac{r^2}{2l^2} \left ( \frac{Q_1'}{Q_1} + \frac{Q_5'}{Q_5} \right ), \nn
\eea
where again $\tilde{g} \ll 1$. 

In \cite{Kanitscheider:2006zf} the field theory deformation corresponding to \eqref{harm2} was shown to be 
\be
I \rightarrow I +  \frac{1}{l^2} \int d^2w  \left ( (\frac{Q_1'}{Q_1} + \frac{Q_5'}{Q_1'}) {\cal O}_{\tau_0}  + (\frac{Q_1'}{Q_1} - \frac{Q_5'}{Q_5}) {\cal O}_{t_0} \right ) + \cdots
\ee
where $({\cal O}_{\tau_0}, {\cal O}_{t_0})$ are dimension four operators, the top components of short multiplets generated from dimension two chiral primaries through the action of the supercharges. These dimension two chiral primaries can be expressed in terms of the fields of \eqref{D1D5cft} as follows. The dimension two primary associated with the $(2,2)$ cohomology is in the untwisted sector, 
\be
\Psi_A^1 (z) \Psi_A^{2 \dagger}(z) \tilde{\Psi}^1_A (\bar{z}) \tilde{\Psi}^{2 \dagger}_A (\bar{z})
\ee
where
\be
\Psi_A^1 = \psi_A^1 + i \psi_A^2; \qquad
\Psi_A^2 = \psi_A^3 + i \psi_A^4; \qquad
\tilde{\Psi}_A^1 = \tilde{\psi}_A^1 + i \tilde{\psi}_A^2; \qquad
\tilde{\Psi}_A^2 = \tilde{\psi}_A^3 + i \tilde{\psi}_A^4.
\ee
The dimension two primary associated with the $(0,0)$ cohomology is in the twist three sector, $\Sigma^{(2)}(z,\bar{z})$, see
\cite{David:2002wn} for more details. 

When each brane stack has the same fraction of D1-branes as D5-branes, $Q_1'/Q_1=Q_5'/Q_5$, only one of the two operators is sourced. The effective description of the inner throat region around one brane stack is via three-dimensional gravity coupled to the scalar field dual to 
${\cal O}_{\tau_0}$, i.e. 
\be
I = \frac{1}{16 \pi G_3} \int d^3 x \sqrt{-g} \left ( {\cal R} + 2 - \frac{1}{2} ( (\partial \tau_0)^2 + m_{\tau_0}^2 \tau_0^2) + \cdots \right )
\ee 
where $m_{\tau_0}^2 = 8$. Here the ellipses again denote contributions from additional Kaluza-Klein modes, associated with subleading terms in \eqref{harm2}.

\subsection{AdS Reissner-Nordstr\"{o}m}

AdS Reissner-Nordstr\"{o}m black holes have received considerable attention in the AdS/CMT literature. The action is Einstein-Maxwell with cosmological constant
\be
I = \frac{1}{16 \pi G_{d+1}} \int d^{d+1} x \sqrt{-g} \left ( {\cal R} + d (d-1) - \frac{1}{4}F^2 \right )
\ee
for which the Einstein equation is 
\be
R_{mn} = - d g_{mn} + \frac{1}{2} F_{mp} F_{n}^{\;p} + \frac{1}{8 (1-d)} F^2 g_{mn}
\ee
and the AdS-RN solution is
\bea
ds^2 &=& \frac{1}{z^2} \left ( - h(z) dt^2 + \frac{dz^2}{h(z)} + dx \cdot dx \right ) \label{rn} \\ 
A &=& \frac{\mu}{z_o^{2-d}} \left (1 - \left (\frac{z}{z_o} \right )^{d-2} \right ) dt \nn
\eea
with
\be
h(z) = 1 - m z^d + \frac{\mu^2}{\gamma^2} z^{2 (d-1)}; \qquad \gamma^2 = \frac{2(d-1)}{(d-2)},
\ee
and implicitly we take $d > 2$. The horizon is at $z = z_o$ and the parameters $m$ and $\mu$ can be expressed in terms of $z_o$ in the extremal solution as follows
\be
\mu^2 = \frac{2 d (d-1)}{(d-2)^2} z_o^{2(1-d)}; \qquad 
m = \frac{2 (d-1)}{(d-2)} z_o^{-d}. 
\ee
from which one can show that 
\be
h''(z_o) = \frac{2d (d-1)}{z_o^2}; \qquad h^{(3)}(z_o) = 2d (5 - 8d + 3d^2) z_o^{-3}.
\ee
Now define $z = z_o + \rho$. In the near horizon limit of the extremal solution
\be
h(\rho) = \frac{d (d-1)}{z_o^2} \rho^2 + \frac{d (5 - 8d + 3d^2)}{3 z_o^3} \rho^3 + \cdots 
\ee
and the leading order metric and potential are 
\bea
ds^2 &=&  \left ( - \frac{d(d-1)}{z_0^4} \rho^2 dt^2 + \frac{d \rho^2}{d (d-1) \rho^2} \right ) + \frac{1}{z_o^2} dx \cdot dx; \label{spec1} \\
A_t &=& - \sqrt{2d (d-1)} \frac{\rho}{z_o^2}. \nn 
\eea
Here the $AdS_2$ curvature radius $l^2$ is such that $l^2 = 1/d (d-1)$
and the  transverse space is flat, due to the cancellation between the cosmological constant and the gauge field contributions. It is convenient to rescale so that 
\bea
ds^2 &=& \frac{1}{d(d-1)}  \left ( - \rho^2 d\td{t}^2 + \frac{d \rho^2}{ \rho^2} \right ) + \frac{1}{z_o^2} dx \cdot dx; \\
A_{\tilde{t}} &=& -  \frac{\sqrt{2}}{\sqrt{d(d-1)}}  \rho. \nn 
\eea
In the AdS/CMT literature this near horizon limit is often interpreted as being dual to a one-dimensional CFT (i.e. a chiral CFT or conformal quantum mechanics). Note that for this interpretation to be valid the transverse space must be compactified so that its spectrum is discrete. 

As in the previous sections, the effects of the outside region can be seen as an irrelevant deformation of the low energy theory. Indeed, we will now show that the deformation is the exact analogue of the deformation involved in the case of the D3-branes.
The chiral operator content of the theory dual to \eqref{spec1} is obtained by diagonalising the linearised equations of motion. Here we do not need to obtain the full spectrum but we will simply focus on the operators of interest. Let the background metric and gauge field be denoted $\bar{g}_{mn}$ and $\bar{A}_{m}$ respectively and perturb the fields as $g_{mn} = \bar{g}_{mn} + h_{mn}$ and $A_{m} = \bar{A}_{m} + a_{m}$. These fields can be decomposed in terms of harmonics of the transverse space:
\bea
h_{\mu \nu} &=& h_{\mu \nu}^I Y^I(x); \qquad h_{\mu i} =  b^{I_v}_{\mu} Y^{I_v}_{i}; \\
h_{ij} &=& \frac{1}{(d-1)}  \pi^I Y^I(x) \bar{g}_{ij} + \phi^{I_t} Y^{I_t}_{(ij)}; \nn \\
a_{\mu} &=& a_{\mu}^I Y^I(x); \qquad a_{i} = a^{I_v} Y_i^{I_v}. \nn
\eea
Here we impose a de Donder gauge $\nabla^{m} h_{mn} = \nabla^{m} a_{m} = 0$. The fields are expressed in terms of eigenmodes of the compact space: $Y^I$ are scalar eigenmodes; $Y^{I_v}_i$ are vector eigenmodes and $Y^{I_t}_{(ij)}$ are traceless symmetric tensor modes. At the linear level the fields $(h_{\mu \nu}^I, \pi^I, a^{I}_{\mu})$ can mix with each other, as can $(b_{\mu}^{I_v},a^{I_v})$. The mode $\phi^{I_t}$ is necessarily decoupled. 

Let us focus on the zero mode of the transverse space, i.e. that associated with the trivial scalar harmonic $Y = 1$: only $(h^{0}_{\mu \nu}, \pi^0,a_{\mu}^0)$ need to be switched on. The independent equations are the $(\mu \nu)$ Einstein equations, the trace of the $(ij)$ Einstein equations and the $\mu$ component of the gauge field equation. 
It is straightforward to show that the field $\pi^0$ satisfies a field equation
\be
\Box \pi^0 = m_{\pi}^2 \pi^0
\ee
with $m_{\pi}^2 = 2/l^2$. The other fields are then determined in terms of $\pi^0$. The dimension of the operator dual to $\pi^0$ is two, which is an irrelevant operator; the dimension is twice the spacetime dimension of the dual CFT, as in the previous examples. In the explicit solution \eqref{rn} we can read off $\pi^0 \sim \rho$, which is indeed consistent with a source term for this operator. Therefore, once again the effective description of the inner throat region is gravity coupled to a scalar field dual to an irrelevant operator. 

\section{Prototype holographic description} \label{sec:six}

In all examples in sections \ref{d3} and \ref{sec:five} the inner throat region can be described in terms of an effective theory consisting of Einstein gravity coupled to massive scalar fields. Therefore a generic model which should suffice to capture the holographic description of field space entanglement entropy is the following:
\be
I = \frac{1}{16 \pi G_{d+1}} \int d^{d+1}x \sqrt{-g} \left  ({\cal R}  + d (d-1) - \frac{1}{2}  ( (\partial T)^2 + m^2 T^2) \right ) \label{fe1}
\ee
where the mass of the scalar field $T$ is is such that $T$ is dual to an irrelevant operator, i.e. $m^2 > 0$. 
This action agrees to leading order with those derived in the previous sections, with the operators being of dimension $2d$,  although the actions in previous sections receive corrections from other higher mass Kaluza-Klein fields (higher dimension operators). 

Let us take the following ansatz for the metric and scalar field
\be
ds^2 = g(r) \frac{dr^2}{r^2} +  r^2 {f(r)} dx \cdot dx; \qquad
T = T(r) \label{abv}
\ee
and work perturbatively in the field $T$ around the AdS background, assuming a non-normalizable mode for $T$. Therefore the leading order solution is an AdS metric with 
\be
T  =   \left ( \frac{r}{l} \right )^{\Delta - d} 
\ee
where we assume a cutoff scale $r_c$ such that $r_c \ll l$. Working perturbatively in $r/l$ one can compute the backreaction on the metric using the Einstein equations: the gauge invariant combination
\be
g(r) - r \partial_r f(r) = 1 + \frac{ (\Delta - d)}{2 (d-1)} \left ( \frac{r}{l} \right )^{2 (\Delta - d)}  + \cdots \label{abv2}
\ee
In a gauge in which $g(r) = 0$
\be
f(r) = 1 - \frac{1}{4 (d-1)} \left ( \frac{r}{l} \right )^{2 (\Delta - d)} + \cdots 
\ee
This model captures the essential features from the systems discussed in the previous sections. We interpret the non-normalizable mode for $T$ in terms of irrelevant deformations of the CFT, obtained by integrating out other fields. Now let us consider the definition of field space entanglement entropy in such a holographic setup. The new functional proposed by  \cite{Mollabashi:2014qfa} is not relevant, since we have no compact part of the geometry. The definition of field space entanglement entropy should however only require a throat region, and this is indeed exactly what is captured by \eqref{abv}. 

Any definition of a holographic functional for the field space entanglement entropy should satisfy the following properties:
\begin{enumerate}
\item The functional should vanish when evaluated on $AdS_{d+1}$ spacetimes, since such backgrounds describe the ground state in a dual CFT which is not entangled with any other field theory.
\item The functional should also vanish when evaluated on 
static, asymptotically locally $AdS$ spacetimes with no horizons for the same reason: such backgrounds describe pure states in (relevantly deformed) CFTs which are not entangled with any other field theory. 
\item The functional should give a non-vanishing result for a spacetime whose asymptotics correspond to an irrelevantly deformed field theory. 
\item The functional should generically give rise to an entanglement entropy of the form $S^F \propto c V_{d-1} \Lambda^{d-1}$ where $V_{d-1}$ is the volume of the spatial sections, $\Lambda$ is the UV cutoff and $c$ is a dimensionless coupling, parameterising the interactions. 
\end{enumerate}
For simplicity, we restrict to static situations. Let us now assume that the holographic functional depends only on the Einstein geometry, as the Ryu-Takanagi functional does; in other words, the functional should not depend on other matter in the bulk theory. Then the simplest possibilities meeting the above requirements are (i)  the area of a spatial cutoff surface and (ii) the spatial volume.

\subsection{Differential entropy} \label{six-de}

First we consider  the area of a spatial cutoff surface of the inner throat and its
relation to field space entanglement entropy and differential entropy. 
The area of such a surface is a natural candidate for an entanglement entropy; 
it has been conjectured  in \cite{{Bianchi:2012ev}} that the entanglement entropy between the degrees of freedom in any given spacetime region and those of its complement is given by the black hole formula to leading order (whenever the leading low energy effective gravitational action is Einstein-Hilbert):
\be \label{hole}
{\cal E} = \frac{ {\cal A}}{4 G_{d+1}}. 
\ee 
Moreover the quantity ${\cal E}$ has a precise definition in terms of field theoretic quantities, when computed in the bottom up system \eqref{fe1}: it is the differential entropy, defined as \cite{Balasubramanian:2013rqa,Balasubramanian:2013lsa,Myers:2014jia,Czech:2014wka,Balasubramanian:2014sra}
\be
{\cal E} = \sum_{k=1}^{\infty} \left [ S(I_k) - S(I_k \cap I_{k+1}) \right ]
\ee
where $\{ I_k \}$ is a set of intervals that partitions the boundary and 
$S(I)$ is the standard entanglement entropy, computed holographically using the Ryu-Takayanagi formula. 
In this case we cover the boundary with $n$ intersecting slabs each of width $\Delta w$ such that the overlap is $(\Delta w -L_w/n)$ where $L_x$ is the regularised length of one of the spatial directions. We then take the limit $n \rightarrow \infty$. 

We now show explicitly that the differential entropy computes the area of a hole in the throat geometries, extending the work of  \cite{Balasubramanian:2013rqa,Balasubramanian:2013lsa,Myers:2014jia,Czech:2014wka,Balasubramanian:2014sra} to cases in which the field theory is deformed by irrelevant deformations (i.e. non AdS asymptotics). 
We parameterise the metric as 
\be
ds^2 = \frac{g(z) dz^2}{z^2} + \frac{f(z) dx \cdot dx}{z^2}
\ee
where the gauge invariant combination $(g(z) + z f'(z))$ is determined by the Einstein equation. Here it is convenient to fix a gauge in which $f(z) = 1$ and hence
\be
g(z) = 1 + \frac{\lambda}{z^{2d}},
\ee
with $\lambda$ a constant, which according to \eqref{abv2} is given by 
\be
\lambda = \frac{d}{2 (d-1) l^{2d}}. \label{l-def}
\ee
Slab entangling regions are minimal surfaces for the functional
\be
S = \frac{1}{4 G_{d+1}} \int dz d^{d-2} x \frac{1}{z^{d-1}} \sqrt{(g(z) + (\partial_z w)^2)},
\ee
where the boundary entangling region extends over $(d-2)$ spatial coordinates but partitions the $w$ direction. From this functional we can immediately write a first integral
\be
\partial_z w = \frac{c g(z)^{1/2} z^{d-1}}{(1 - c^2 z^{2 (d-1)})^{\frac{1}{2}}},
\ee
where $c$ is an integration constant, relating to the turning point $z_{\ast}$ of the minimal surface:
\be
c z_{\ast}^{d-1} = 1
\ee
Thus
\be
\Delta w = z_{\ast} \int^{1}_{\epsilon/z_{\ast}} \frac{g(z)^{1/2} s^{d-1}}{\sqrt{1 - s^{2(d-1)}}} ds
\ee
and
\be
S  = \frac{V_{d-2}}{4 z_{\ast}^{d-2} G_{d+1}} \int^1_{\epsilon/z_{\ast}} \frac{g(z)^{1/2} } {s^{d-1} \sqrt{1 - s^{2(d-1)}}}ds, 
\ee
where $V_{d-2}$ is the regulated volume of the $(d-2)$ spatial coordinates. 

Let us parameterise the regulated elliptic integrals as follows:
\bea
{\cal L}_{\epsilon \rightarrow 0} \left ( \int^{1}_{\epsilon/z_{\ast}} \frac{s^{d-1}}{\sqrt{1 - s^{2(d-1)}}} ds \right ) &=& k_1; \\
{\cal L}_{\epsilon \rightarrow 0} \left ( \frac{1}{z_{\ast}^{2d}} \int^{1}_{\epsilon/z_{\ast}} \frac{1}{s^{d+1} \sqrt{1 - s^{2(d-1)}}} ds \right ) &=&  \frac{k_2}{z_{\ast}^d \epsilon^d} + \frac{k_3}{z_{\ast}^{2d}}; \nn \\
{\cal L}_{\epsilon \rightarrow 0} \left (  \frac{1}{z_{\ast}^{d-2}} \int^{1}_{\epsilon/z_{\ast}} \frac{1}{s^{d-1}\sqrt{1 - s^{2(d-1)}}} ds \right ) &=& \frac{K_1}{\epsilon^{d-2}} + \frac{K_2}{z_{\ast}^{d-2}}; \nn \\
{\cal L}_{\epsilon \rightarrow 0} \left (  \frac{1}{z_{\ast}^{3d-2}} \int^{1}_{\epsilon/z_{\ast}} \frac{1}{s^{3d-1}\sqrt{1 - s^{2(d-1)}}} ds \right ) &=& \frac{K_3}{\epsilon^{3d-2}} + \frac{K_4}{\epsilon^d z_{\ast}^{2 (d-1)}} + \frac{K_5}{z_{\ast}^{3d-2}}. \nn
\eea

Then the entanglement entropy is given by
\be
S =  \frac{V_{d-2}}{4 G_{d+1}}  \left ( \frac{K_1}{\epsilon^{d-2}} + \frac{K_2}{z_{\ast}^{d-2}} + \frac{\lambda}{2} \left  ( \frac{K_3}{\epsilon^{3d-2}} + \frac{K_4}{\epsilon^d z_{\ast}^{2 (d-1)}} + \frac{K_5}{z_{\ast}^{3d-2}} \right ) \right ), 
\ee
while
\be
\Delta w = z_{\ast} \left ( k_1 + \frac{\lambda}{2} \left ( \frac{k_2}{z_{\ast}^d \epsilon^d} + \frac{k_3}{z_{\ast}^{2d}} \right )  \right ) \label{width}
\ee
The differential entropy is given by
\be
{\cal E} = L_w \frac{\partial z_{\ast}}{\partial (\Delta w)} \frac{\partial S}{\partial z_{\ast}}, 
\ee
where $L_w$ is the regulated length of the $w$ direction. This expression evaluates to give the area of a hole, i.e. the differential entropy is
\be
{\cal E } = \frac{V_{d-1}}{4 z_{\ast}^{d-1} G_{d+1}} \left (1 + {\cal O} (\lambda^2) \right ), \label{ads-r}
\ee
with $V_{d-1}$ the regulated volume of all spatial directions, provided that
\be
k_1 = - (d-2) K_2; \qquad
k_2 =  2 K_4; \qquad
(2d-1) k_3 = (3d-2) K_5.  
\ee
These identities are indeed satisfied with
\be
k_1 = \sqrt{\pi} \frac{\Gamma ( \frac{d}{2(d-1)} )}{\Gamma( \frac{1}{2 (d-1)} )}; \qquad
k_2 = \frac{1}{d}; \qquad
k_3 =  \frac{\sqrt{\pi}}{2(d-1)} \frac{\Gamma (- \frac{d}{2(d-1)} )}{\Gamma( -\frac{1}{2 (d-1)} )}. 
\ee
For completeness note that
\be
K_1 = \frac{1}{(d-2)}; \qquad K_3 = \frac{1}{(3d-2)}. 
\ee
It might seem surprising that the relation between the strip width and the depth of the entangling surface \eqref{width} depends on the UV cutoff. However, one can rewrite this relation using \eqref{l-def}
\be
\Delta w = z_{\ast} \left ( k_1 + \frac{d}{4 (d-1) l^d z_{\ast}^d} \left ( k_2 {g}^d + \frac{k_3}{ l^{d} z_{\ast}^{d}} \right )  \right ), \label{width2}
\ee
where ${g} = 1/l \epsilon \ll 1$. Similarly we can rewrite the entanglement entropy for each strip as
\be
S =  \frac{V_{d-2}}{4 G_{d+1}}  \left ( \frac{1}{\epsilon^{d-2}} \left (K_1 + \frac{d}{4(d-1)} {g}^2 K_3 \right )
+ \frac{1}{z_{\ast}^{d-2}} \left (K_2 + \frac{d}{4(d-1)}  (\frac{g K_4}{ l^d z_{\ast}^d} + 
 \frac{K_5}{l^{2d} z_{\ast}^{2d}}   ) \right ) \right ). 
\ee


Note that the quantity \eqref{hole} is inherently dependent on the choice of gauge for the radial coordinate. 
If we compute \eqref{hole} on a surface $r_{\ast}$ in the metric \eqref{abv2} we obtain
\be
{\cal E} = \frac{V_{d-1}}{4 G_{d+1}} {r}_{\ast}^{d-1} \left ( 1 -  \frac{1}{8} \left (\frac{ {r}_{\ast}}{l} \right )^{2 d} \right ). \label{de-2}
\ee
This matches \eqref{ads-r} if we take into account the redefinition of the radial coordinate, i.e. 
\be
\frac{1}{z} = {r} \left (1 - \frac{1}{8 (d-1)} \left (\frac{ r}{l} \right )^{2 d} \right ). 
\ee
The differential entropy evaluated on the cutoff $r_{\ast} = r_c$ gives
\be
{\cal E} = \frac{V_{d-1}}{4 G_{d+1}} {r}_{c}^{d-1} \left ( 1 -  \frac{1}{8} \tilde{g}^{2 d} \right ).
\ee
Clearly if we define $\delta {\cal E}$ as the difference between the differential entropy in AdS and that in the deformed background then 
\be
\delta {\cal E} = - \tilde{g}^{2d} \frac{V_{d-1}}{32 G_{d+1}} {r}_{c}^{d-1}, \label{deltae}
\ee
which is of the same form as the field space entanglement entropy. 
We can understand the physical difference between \eqref{ads-r} and \eqref{de-2} as follows. In \eqref{ads-r} we effectively adjust the width of the slabs used to subdivide the space in the field theory to ensure that the area of the bulk hole remains unchanged at first order, i.e. to enforce the bulk gauge in which $f(z) = 1$, while in \eqref{de-2} we allow the area of the bulk hole to be changed. The most natural coordinate gauge choice from the field theory perspective is however neither of these: from the field theory one would usually fix the strip width $\Delta w$ and compute the area of the associated hole.

\subsection{Spatial volume of inner throat}

Let us now consider the spatial volume. It might seem surprising to define a measure of entanglement in terms of a spatial volume but one can give a heuristic argument in favour of this possibility as follows. For the usual entanglement entropy, degrees of freedom are entangled at the surface separating spatial regions A and B; the extension of this surface into the bulk gives the Ryu-Takayanagi minimal surface. In the case at hand, degrees of freedom have been integrated out throughout the whole spatial region; the image of this region in the bulk is the entire spatial volume. 

However, the naive spatial volume clearly violates the first and second requirements stated above, as one necessarily obtains a non-zero answer for $AdS$ and asymptotically $AdS$ throats. To try to solve this problem, one could use the renormalised spatial volume:
\be
S^V = \frac{C}{G_{d+1}} \int_{\Sigma} d^{d} x \sqrt{\gamma} - \frac{C}{(d-1) G_{d+1}} \int_{\partial \Sigma} d^{d-1} x \sqrt{h} \left (1 - \frac{1}{ 2(d-2) (d-3)} {\cal R}(h) + \cdots \right ) \label{func}
\ee
where $\Sigma$ is a hypersurface of constant time in the $(d+1)$-dimensional stationary manifold and $\partial \Sigma$ is its boundary. Here $C$ is an overall normalisation. The Ricci scalar of the boundary metric is denoted ${\cal R}(h)$ and the boundary counterterms renormalise the volume. To work out these counterterms we use the following expansions of a static, asymptotically locally $AdS_{d+1}$ Einstein manifold in Fefferman-Graham coordinates near the conformal boundary \cite{deHaro:2000xn}
\bea
ds^2 &=& \frac{dz^2}{z^2} + \frac{1}{z^2} \left ( g_{tt}  dt^2 + g_{ij} dx^i dx^j \right ); \label{fg} \\
g_{tt} &=& - \left  (1 + \frac{z^2}{2(d-1)}  {\cal R}(g_{(0)})+ \cdots \right  ); \nn \\
g_{ij} &=& g_{(0) ij}  + z^2 \frac{1}{(d-2)} \left ( - {\cal R}_{ij} (g_{(0)}) + \frac{1}{2(d-1)} {\cal R}(g_{(0)}) g_{(0) ij}  + \cdots \right ). \nn
\eea
Following \cite{Graham:1999pm,Graham:1999jg}, we see that the volume term gives logarithms in even bulk dimensions, which must be cancelled by logarithmic counterterms. In odd bulk dimensions ($d$ even) there are no finite counterterms and no finite contributions from the counterterms. In even bulk dimensions ($d$ odd) finite counterterms can be included. 

The first example of logarithmic divergences arises in $d=3$. In this case the action above becomes
\be
S^V = \frac{C}{G_{d+1}} \int_{\Sigma} d^{3} x \sqrt{\gamma} - \frac{C}{2 G_{d+1}} \int_{\partial \Sigma} d^{2} x \sqrt{h} \left (1 + \frac{1}{2} {\cal R}(h) \ln \epsilon + \alpha_{s} {\cal R}(h)  \right ),
\ee
where the regulating surface is at $z = \epsilon$. (For $d=3$ the above action is not-defined, because of the logarithmic term.) Here the choice of $\alpha_s$ determines the scheme, since this counterterm is finite. 

By construction this functional vanishes for a constant time slice of $AdS_4$ in Poincar\'{e} coordinates. However, it is never possible to fix finite counterterms such that the functional vanishes for all asymptotically locally AdS solutions with no horizons since finite contributions depend not only on the non-normalizable data of the metric, but also on the normalizable data. In particular, even for a constant time slice of $AdS_4$ in global coordinates $S^V$ is finite but non-zero:
\be
S^V = \frac{4 \pi C}{G_{d+1}} \left ( \frac{1}{4} - \frac{1}{2} \ln 2 - \alpha_s \right ).
\ee
One can fix $\alpha_s$ to set this to zero, but one will then still obtain finite answers for other conformal classes of the boundary metric. 

Another conceptual issue with the use of the renormalised volume is that 
the Fefferman-Graham expansion for the metric changes as the matter content is adjusted. Counterterms to renormalise the volume therefore depend not just on the induced geometry but also on the boundary values of the matter fields, so the functional does not in general depend only on the geometry. 


Leaving these issues to one side for a moment, the functional can be computed for the solution \eqref{abv}, working perturbatively in the irrelevant field. We set
\be
g(r) = 1 + \Delta g(r); \qquad f(r) = 1 + \Delta f(r),
\ee
where the gauge invariant combination $(\Delta g(r) - r \partial \Delta f(r))$ is determined from \eqref{abv2}. 
Then the renormalised spatial volume gives 
\be
S^V  = \frac{C V_{d-1}}{2 G_{d+1}} \int_{0}^{r_c} dr r^{d-2} (\Delta g(r) - r \partial \Delta f(r)) \label{ent-gr}
\ee
where $V_{d-1}$ is again the regulated volume of the spatial sections. The integral is expressed in terms of the gauge invariant metric perturbation and evaluating it we obtain  
\be
S^V = C \frac{( \Delta - d) }{4(d-1) ( 2 \Delta - d -1)} \frac{V_{d-1} r_c^{d-1}}{G_{d+1}} \td{g}^{2 ( \Delta -d )}, \label{ent}
\ee
where we define
\be
\frac{r_c}{l} = \td{g}
\ee
The functional therefore by construction does have the same qualitative behaviour as the field space entanglement entropy of section \ref{two}.

\subsection{Summary and interpretation}

In the holographic systems we have been discussing, degrees of freedom of the dual field theory are indeed localised in the inner throat region. In particular, in the Coulomb branch geometries the low energy physics is analogous to that in the field theory examples in section \ref{two}. However, imposing any cutoff on the inner throat region automatically removes not only the degrees of freedom associated with the other brane stacks (as in the field theory examples), but also high energy modes from the $SU(N/2)$ CFT dual to the inner throat (which were not removed by the definition of field theory entanglement entropy of section \ref{two}). 

It is hence unsurprising that both geometric measures of entanglement, the spatial volume of the inner throat and the area of the cutoff of this throat, give non-zero answers even for asymptotically AdS throats. The latter are dual to states in a single conformal field theory and, according to the field theory definition, their field space entanglement entropy would be zero. Yet the geometric entanglement quantities both remove high energy modes from this CFT by introducing a geometric cutoff, resulting in non-zero entanglement.  The procedures of background subtraction \eqref{deltae} and renormalisation \eqref{func} subtract off the entanglement with high energy modes of the CFT. 

Corrections to the throat geometry associated with irrelevant deformations give analogous contributions to the geometric entropies to the field space entanglement entropy of section \ref{two}. The measure of entanglement associated with the spatial volume of the throat has the advantage that it is gauge invariant, see
\eqref{ent-gr}, but the exact dual field theory definition is unclear. The differential entropy by contrast has a sharp field theory definition although the relation between the defining strip width in the quantum field theory and the cutoff surface is subtle. 

Finally, let us note that both geometric quantities are also qualitatively similar to the momentum space entanglement entropy explored in \cite{Balasubramanian:2011wt}, although the difficulty in exactly matching the bulk radial coordinate with boundary RG scale precludes a precise identification.

\section{Conclusions} \label{sec:conc}

In this paper we have derived effective descriptions of inner throat regions in holographic geometries in terms of Einstein gravity coupled to massive scalar fields dual to irrelevant operators. Such descriptions are applicable both to Coulomb branch solutions with separated brane stacks and to the interior region of near extremal anti-de Sitter black branes. 

Using these effective descriptions we have explored geometric measures of entanglement, characterising the entanglement of the degrees of freedom associated with the inner throat with those degrees of freedom associated with the geometric complement. We showed that the differential entropy computes the area of a hole, even in an irrelevantly deformed conformal field theory. Both the differential entropy and the spatial volume of the throat capture features of the field space entanglement entropy. However, unsurprisingly, the geometric measures of entanglement receive contributions associated with the entanglement with high energy modes of the low energy CFT dual to the inner throat and thus we conclude that the field space entanglement entropy cannot be precisely realised holographically.

The generalized holographic entanglement entropy (associated with a partitioning of the compact part of the bulk geometry) should correspond to a global symmetry entanglement entropy. In future work it would be interesting to sharpen the field theory definition of the latter and compute it in interacting field theories. One would also like to understand whether the generalized holographic entanglement entropy can be understood and derived using the methods of \cite{Lewkowycz:2013nqa}. 

More generally, we note that it is hard to derive precise relationships between bulk (geometric) measures of entanglement and quantum field theory measures of entanglement. From the quantum field theory perspective it would  also be natural to consider entanglement between different components of the matrix for matrix valued fields (although a gauge invariant definition would clearly be needed). From the bulk perspective such entanglement would presumably be hard to realise geometrically for the same reasons discussed in this paper: there is no reason why the degrees of freedom traced out should generically be localized anywhere in the bulk spacetime.

\section*{Acknowledgments}

We would like to thank Da-Wei Pang for collaboration in the early phases of this project, and Ali Mollabashi and Mohammad Mozaffar for sharing unpublished results on field space entanglement entropy. We thank Vijay Balasubramanian and Tadashi Takayanagi for helpful comments and discussions. 
We acknowledge support from a grant of the John Templeton Foundation. The opinions expressed in this publication are those of the authors and do not necessarily reflect the views of the John Templeton Foundation. 
This work was supported by the Science and Technology Facilities Council (Consolidated Grant ``Exploring the Limits of the Standard Model and Beyond'') and by the Engineering and Physical Sciences Research Council. 
This work was supported in part by National Science Foundation Grant No. PHYS-1066293 and the hospitality of the Aspen Center for Physics.
We thank the Galileo Galilei Institute for Theoretical Physics for the hospitality and the INFN for partial support during the completion of this work.

\appendix

\section{M2-brane analysis} \label{appa}

In this section we discuss the Kaluza-Klein spectrum on $AdS_4 \times S^7$, focussing on the scalar fields of interest dual to irrelevant operators. Our discussion is consistent with the early papers \cite{Biran:1983iy,Casher:1984ym} but  we follow the approach introduced in \cite{Kim:1985ez} for $AdS_5 \times S^5$. 

The Einstein equation is
\be
R_{MN}  = \frac{1}{6} F_{MPQR} F_{N}^{\; \; PQR} -  \frac{1}{72} F_{PQRS} F^{PQRS} g_{MN},
\ee
while the equation for the four-form is
\be
D_{M} F^{MNPQ} = \frac{1}{\sqrt{2} 576} \eta^{M_1 \cdots M_8 NPQ} F_{M_1 \cdots M_4} F_{M_5 \cdots M_8},
\ee
with $F_{MNPQ} = 24 \partial_{[M} A_{NPQ ]}$. 

For $AdS_4 \times S^7$ the background solution can be expressed as 
\bea
R_{\alpha \beta \gamma \delta} &=& m_7^2 (g^o_{\alpha \gamma} g^o_{\beta \delta} - g^o_{\alpha \delta} g^o_{\beta \gamma}); \\
R_{\mu \nu \rho \sigma} &=& - m_4^2 (g^o_{\mu \rho} g^o_{\nu \sigma} - g^o_{\mu \sigma} g^o_{\mu \rho}), \nn
\eea
where $m_4^2 = 4 m_7^2$ and 
\be
F^o_{\mu \nu \rho \sigma} = \sqrt{18} m_7 \eta_{\mu \nu \rho \sigma}. 
\ee
Now we consider metric perturbations such that 
\be
g_{MN} = g^o_{MN} + h_{MN}.
\ee
Fixing a gauge such that 
\be
D^{\alpha} h_{\alpha \beta} = D^{\alpha} h_{\alpha \mu} = 0, 
\ee
with $D^{\alpha}$ being the background covariant derivative, 
we can decompose the metric perturbations in terms of spherical harmonics as 
\bea
h_{\mu \nu} &=& \sum h_{\mu \nu}^I(x) Y^I(y); \qquad h_{\mu \alpha} = \sum B_{\mu}^{I_v}(x) Y_{\alpha}^{I_v}(y); \label{appa7} \\
h_{(\alpha \beta)} &=& \sum \phi^{I_t} (x) Y_{(\alpha \beta)}^{I_t}(y); \qquad
h^{\alpha}_{\alpha} = \sum \pi^I(x) Y^I(y). \nn
\eea
Here $Y^I(y)$ denote scalar harmonics; $Y_{\alpha}^{I_v}$ denote vector harmonics which satisfy $D^{\alpha} Y_{\alpha}^{I_v} = 0$ and $Y_{(\alpha \beta )}$ denote tensor harmonics satisfying $D^{\alpha} Y_{(\alpha \beta)} = 0$.

Similarly the three form can be expressed as $A_{MNP} = A^{o}_{MNP} + a_{MNP}$ with
\bea
a_{\mu \nu \rho} &=& \sum a_{\mu \nu \rho}^I Y^I; \qquad
a_{\mu \nu \alpha} = \sum a_{\mu \nu}^{I_v} Y^{I_v}_{\alpha}; \\
a_{\mu \alpha \beta} &=& \sum a_{\mu}^{I_a} Y^{I_a}_{[ \alpha \beta ]}; \qquad
a_{\alpha \beta \gamma} = \sum a^{I_3} Y^{I_3}_{[ \alpha \beta \gamma ]},
\eea
where we have imposed a gauge choice:
\be
D^{\alpha} a_{\alpha\mu \nu} = D^{\alpha} a_{\alpha \beta \mu} = D^{\alpha} a_{\alpha \beta \gamma} = 0. 
\ee
Here $Y_{[\alpha \beta ]}^{I_a}$ and $Y^{I_3}_{[\alpha \beta \gamma]}$ are again tensor harmonics. 

The linearized Einstein and four-form equations can be diagonalised by projecting onto the linearly independent spherical harmonic components. Clearly only modes associated with the same spherical harmonics can mix at linear order. Thus in particular we need to diagonalize the equations for $(h_{\mu \nu}^I, \pi^I,a_{\mu \nu \rho}^I)$. Note that we can dualise
$a_{\mu \nu \rho}^I$ and express it as
\be
a_{\mu \nu \rho}^I = \eta^o_{\mu \nu \rho \sigma} t^{I \sigma}. 
\ee 
To diagonalize the equations of motion we need the projections of the $(\mu \nu)$ Einstein equation,  the $(\alpha \beta)$ Einstein equation and the four form equation onto scalar harmonics. Projecting the symmetric traceless part of the $(\alpha \beta)$ Einstein equation we obtain
\be
\left ( h_{\mu}^{\mu I} + \frac{5}{7} \pi^I \right ) D_{(\alpha} D_{\beta )} Y^I = 0,
\ee
which immediately allows $h_{\mu}^{\mu I}$ to be eliminated in favour of $\pi^I$. From the four form equation we obtain 
\be
\left ( t^{\rho I}_{; \rho \mu} + \Box_y t_{\mu}^I - \frac{1}{2} \sqrt{18} m_7 (h^{\rho I}_{\rho} - \pi^I)_{; \mu} \right ) Y^I = 0,
\ee
where $\Box_y$ is the Laplacian along the $S^7$. In particular, for modes depending on the trivial constant spherical harmonic:
\be
\left ( t^{\rho 0}_{; \rho} + \frac{6 \sqrt{18}}{7} m_7 \pi^0 \right )_{;\mu} = 0.
\ee
Tracing the $(\alpha \beta)$ Einstein equation with $g^{o \alpha \beta}$ and projecting onto the trivial spherical harmonic
then gives the following equation of motion for $\pi^0$: 
\be
\left ( \Box_x \pi^0  - 72 m_7^2 \pi^0 \right ) = \left ( \Box_x \pi^0 - 18 m_4^2 \pi^0 \right ) = 0, \label{appa15}
\ee 
where we have eliminated $t^{\rho 0}_{; \rho}$ and $h^{\mu 0}_{\mu}$ using the relations above.  The mass implies that the field 
$\pi^0$ is dual to an operator of dimension six.

\section{M5-brane analysis} \label{appb}

In this section we discuss the Kaluza-Klein spectrum on $AdS_7 \times S^4$, focussing on the scalar fields of interest dual to irrelevant operators. Our discussion is consistent with the results of \cite{Pilch:1984xy,Gunaydin:1984wc,vanNieuwenhuizen:1984iz} but we follow the elegant approach introduced in \cite{Kim:1985ez} for $AdS_5 \times S^5$.

For $AdS_7 \times S^4$ the background solution can be expressed as 
\bea
R_{\alpha \beta \gamma \delta} &=& m_4^2 (g^o_{\alpha \gamma} g^o_{\beta \delta} - g^o_{\alpha \delta} g^o_{\beta \gamma}); \\
R_{\mu \nu \rho \sigma} &=& - m_7^2 (g^o_{\mu \rho} g^o_{\nu \sigma} - g^o_{\mu \sigma} g^o_{\mu \rho}), \nn
\eea
where $\mu$ denotes $AdS$ indices and $\alpha$ denotes $S^4$ indices. Here $m_4^2 = 4 m_7^2$ and 
\be
F^o_{\a \b \gamma \delta} = \sqrt{18} m_7 \eta_{\alpha \beta \gamma \delta}. 
\ee
Now we consider metric perturbations such that 
\be
g_{MN} = g^o_{MN} + h_{MN}.
\ee
Fixing a gauge such that 
\be
D^{\alpha} h_{\alpha \beta} = D^{\alpha} h_{\alpha \mu} = 0 
\ee
we can decompose the metric perturbations in terms of spherical harmonics as 
\bea
h_{\mu \nu} &=& \sum h_{\mu \nu}^I(x) Y^I(y); \qquad h_{\mu \alpha} = \sum B_{\mu}^{I_v}(x) Y_{\alpha}^{I_v}(y); \label{appb5} \\
h_{(\alpha \beta)} &=& \sum \phi^{I_t} (x) Y_{(\alpha \beta)}^{I_t}(y); \qquad
h^{\alpha}_{\alpha} = \sum \pi^I(x) Y^I(y). \nn
\eea
As in the previous section $Y^I$ denote scalar harmonics, $Y_{\alpha}^{I_v}$ denote vector harmonics and $Y^{I_t}_{(\alpha \beta)}$ denote tensor harmonics. 

Similarly the three form can be expressed as $A_{MNP} = A^{o}_{MNP} + a_{MNP}$ with
\bea
a_{\mu \nu \rho} &=& \sum a_{\mu \nu \rho}^I Y^I; \qquad
a_{\mu \nu \alpha} = \sum a_{\mu \nu}^{I_v} Y^{I_v}_{\alpha}; \\
a_{\mu \alpha \beta} &=& \sum a_{\mu}^{I_v} \epsilon_{\alpha \beta \gamma \delta} D^{\gamma} Y^{\delta I_v}; \qquad
a_{\alpha \beta \gamma} = \sum a^{I} \epsilon_{\alpha \beta \gamma \delta} D^{\delta} Y^I,
\eea
where we have imposed a gauge choice:
\be
D^{\alpha} a_{\mu \nu \alpha} = D^{\alpha} a_{\mu \alpha \beta} = D^{\alpha} a_{\alpha \beta \gamma} = 0. 
\ee
The linearized Einstein and four-form equations can be projected into spherical harmonic components and only modes associated with the same spherical 
harmonics can mix at linear order. For the case at hand we only need to diagonalize the equations for $(h_{\mu \nu}^I, \pi^I,a_{\mu \nu \rho}^I, a^I)$. Note that we can dualise
$a_{\mu \nu \rho}^I$ and express it as
\be
a_{\mu \nu \rho}^I = \eta^o_{\mu \nu \rho \sigma} t^{I \sigma}. 
\ee 
To diagonlize the equations of motion we need the projections of the $(\mu \nu)$ Einstein equation,  the $(\alpha \beta)$ Einstein equation and the four form equation onto scalar harmonics. Projecting the symmetric traceless part of the $(\alpha \beta)$ Einstein equation we obtain
\be
\left ( h_{\mu}^{\mu I} + \frac{1}{2} \pi^I \right ) D_{(\alpha} D_{\beta )} Y^I = 0,
\ee
which again immediately allows $h_{\mu}^{\mu I}$ to be eliminated in favour of $\pi^I$. 
Tracing the $(\alpha \beta)$ Einstein equation with $g^{o \alpha \beta}$ and projecting onto the trivial constant spherical harmonic gives
 the following equation for $\pi^0$:
\be
\left (  \Box_x \pi^0  - 72 m_7^2 \pi^0 \right ) =0. \label{appb11}
\ee 
Thus $\pi^0$ is dual to an operator of dimension twelve in the CFT.

\providecommand{\href}[2]{#2}\begingroup\raggedright\endgroup

\end{document}